\documentclass[preprint,3p]{elsarticle}

\usepackage{amssymb}
\usepackage{amsmath}
\usepackage{booktabs} 
\usepackage{graphicx} 
\usepackage{color}
\usepackage{longtable}
\usepackage{pgfplots}
\usepackage{booktabs}
\usepackage{hyperref}
\usepackage{siunitx} 
\usepackage{xspace}
\usepackage{multirow}
\usepackage{makecell}
\pgfplotsset{compat=1.18}
\usepackage[numbers]{natbib}
\usepackage{changepage}
\usepackage{xfp}

\newcommand{\bulletSSS}{{\LARGE $\cdot$}}

\def\CrimeEventsN/{\num{1048}}
\def\DAOEventsN/{\num{33}}
\def\DAOEventsAvailableN/{\num{22}}
\def\CoinGeckoGovTokensN/{\num{110}}
\def\GovOverlapsN/{\num{29}}
\def\LPoolsN/{\num{242054}}
\def\LPoolsGovN/{\num{93}}
\def\UniqueGovAssetsN/{\num{14}}

\def\GovPricesN/{\num{83}}
\def\DirectEcoImpact/{453.9M}
\def\DirectEcoImpactPct/{25.6}
\def\DirectAllEcoImpact/{613.2M}
\def\IndirectEcoImpact/{1.3B}
\def\IndirectEcoImpactN/{\num{12}}
\def\IndirectEcoImpactPct/{74.4}
\def\TotalEcoImpact/{1.8B}
\def\EventExpMinusSixHoursEstimate/{-10.3}
\def\EventExpMinusSixHoursSE/{4.7}
\def\EventExpMinusSixHoursPValue/{\num{0.0715273812163931}}
\def\AllGovDirDamage/{613.2M}
\def\PriceCorAvg/{0.8}
\def\PriceCorSE/{0.0}
\def\VolCorAvg/{1.0}
\def\VolCorSE/{0.0}
\def\PriceCorAvgN/{3.8}
\def\PriceCorSEN/{0.3}
\def\VolCorAvgN/{5.0}
\def\VolCorSEN/{0.2}
\def\SigPriceImpactN/{\num{16}}
\def\SigVolImpactN/{\num{16}}
\def\SigPriceImpactNegN/{\num{12}}
\def\SigPriceImpactNegNPct/{55}
\def\SigPriceImpactPosN/{\num{4}}
\def\SigPriceImpactNegPct/{-13.5}
\def\SigPriceImpactNegMin/{-59.3}
\def\SigPriceImpactNegMax/{-1.5}
\def\SigPriceImpactNegSE/{4.6}
\def\SigVolImpactPosN/{\num{15}}
\def\SigVolImpactPosNPct/{68}
\def\SigVolImpactPosPct/{127.9}
\def\SigVolImpactPosMin/{22.9}
\def\SigVolImpactPosMax/{458.5}
\def\SigVolImpactPosSE/{30.6}
\def\SigPriceTwoDigNegN/{\num{3}}
\def\SigVolThreeDigPosN/{\num{7}}
\def\PriceTrendNeg/{\num{15}}
\def\VolTrendPos/{\num{14}}
\def\CoocSigImp/{\num{7}}
\def\SigIndLargerDirDamPct/{}
\def\MCAvgSig/{110.1M}
\def\SigTotalLoss/{1.3B}
\def\SigTotalWin/{256.7M}
\def\CrimeLossForSigImpact/{453.9M}

\title{The Economic Impact of DeFi Crime Events on Decentralized Autonomous Organizations (DAOs)}

\begin{document}

\begin{frontmatter}

\author[label1,label2]{Stefan Kitzler}
\author[label3]{Masarah Paquet-Clouston}
\author[label1]{Bernhard Haslhofer}

\affiliation[label1]{organization={Complexity Science Hub},
             addressline={Metternichgasse 8},
             city={Vienna},
             postcode={1030},
             country={Austria}}

\affiliation[label2]{organization={AIT Austrian Institute of Technology},
             addressline={Giefinggasse 4},
             city={Vienna},
             postcode={1210},
             country={Austria}}

\affiliation[label3]{organization={Université de Montréal},
	addressline={2900 Edouard Montpetit Blvd},
	city={Montreal},
	postcode={H3T 1J4},
	state={Quebec},
	country={Canada}}

\begin{abstract}
The Decentralized Finance (DeFi) ecosystem has experienced over \$10 billion in direct losses due to crime events.
Beyond these immediate losses, such events often trigger broader market reactions, including price declines, trading activity changes, and reductions in market capitalization.
Decentralized Autonomous Organizations (DAOs) govern DeFi applications through tradable governance assets that function like corporate shares for voting and decision-making.
Leveraging DeFi's granular trading data, we conduct an event study on \DAOEventsAvailableN/ crime events between 2020 and 2022 to assess their economic impact on governance asset prices, trading volumes, and market capitalization.
Using a dynamic difference-in-differences (DiD) framework with counterfactual governance assets, we aim for causal inference of intraday temporal effects.
Our results show that \SigPriceImpactNegNPct/\% of crime events lead to significant negative price impacts, with an average decline of about \num[round-mode=places, round-precision=0]{\fpeval{abs(\SigPriceImpactNegPct/)}}\%. Additionally, \num[round-mode=places, round-precision=0]{\SigVolImpactPosNPct/}\% of crime events lead to increased governance asset trading volume.
Based on these impacts, we estimate indirect economic losses of over \$1.3 billion in DAO market capitalization, far exceeding direct victim costs and accounting for \num[round-mode=places, round-precision=0]{\IndirectEcoImpactPct/}\% of total losses.
Our study provides valuable insights into how crime events shape market dynamics and affect DAOs.
Moreover, our methodological approach is reproducible and applicable beyond DAOs, offering a framework to assess the indirect economic impact on other cryptoassets.

\end{abstract}

\begin{keyword}

Economic Impact \sep Decentralized Finance \sep DeFi \sep Crime \sep Decentralized Autonomous Organization \sep DAO \sep Event Study \sep Difference-in-Difference

\end{keyword}

\end{frontmatter}



\section{Introduction}

In the aftermath of the 2008 financial crisis, a loss of confidence in centralized institutions and financial intermediaries led to a search for alternative financial systems~\cite{Harvey2020}. More recently, Decentralized Finance (DeFi) has emerged as a new financial paradigm, leveraging distributed ledger technology to provide traditional financial services, such as lending, borrowing, or asset exchange, in a decentralized manner~\cite{Auer2024}.
DeFi primarily operates with cryptoassets, also called tokens, which represent digital value or rights within the ecosystem. In this paper, we use the terms ``cryptoassets'' and ``tokens'' interchangeably. As of this writing, approximately \num{60} billion USD in cryptoassets are locked in DeFi services on the Ethereum network.

Over the past years, DeFi has transformed into a distinct economic market with its own technical design~\cite{Werner2022}, dynamics~\cite{Lehar2022}, and governance mechanisms~\cite{Kiayias2023a}.
It operates based on a novel governance structure known as Decentralized Autonomous Organizations (DAOs), which are community-governed and aim to democratize decision-making.
Governance tokens confer voting rights, granting holders the ability to vote on decisions ranging from technical upgrades to economic parameter adjustments~\cite{Liu2021,Kiayias2023}.
These organizations resemble publicly traded companies, as governance decisions are influenced by token holders who exercise voting power over key decisions.

DeFi, as an attractive target for financial criminals, has repeatedly fallen victim to cybercrime. A recent study has shown that cybercrime events on DeFi services directly caused cumulative losses of approximately 10 billion USD by 2022~\cite{CarpentierDesjardins2025}. However, this figure does not account for the broader economic impact of such security incidents. In traditional economies, cybercrime is known to have a larger economic impact going beyond direct costs imposed on targeted organizations and their customers~\cite{Anderson2013,Anderson2019,Agrafiotis2018}. Past research has shown that cybercrime events, such as data breaches or hacks, often trigger adverse market reactions for targeted organizations, leading to negative impacts on their stock prices and market valuations~\cite{Spanos2016,Ebrahimi2022,Tosun2021,Ali2021a,Woods2021}. Hence, the economic consequences of cybercrime are twofold: direct financial costs, such as the amount stolen, and indirect economic costs, such as the loss in market valuation due to reputational damage.

Yet, to date, estimates on the broader indirect economic impact of cybercrime on DAOs are missing in the literature. Assessing how such events impact governance assets cannot only shed light on the financial resilience of DAOs, but also provide insights into how decentralized governance models respond to such events. To close this research gap, this study systematically assesses and quantifies the indirect economic impact of cybercrime, hereafter called DeFi crime events, on associated DAOs. In line with previous research~\cite{Anderson2019,Woods2021,Harvey2020}, this study \textbf{estimates the indirect economic impact of DeFi crime events on prices, trading volumes, and the market capitalization} of governance tokens. To this end, our study leverages new data sources, using granular trading data extracted from the public Ethereum blockchain. The key contributions of this study are as follows:

\begin{enumerate}

	\item We gather a set of \DAOEventsAvailableN/ DeFi crime events affecting \UniqueGovAssetsN/ distinct DAOs and their governance tokens and collect price and trading volume data from Decentralized Exchanges (DEXs) on Ethereum.

	\item We use a dynamic difference-in-differences (DiD) model to measure the temporal effects of DeFi crime events on price, trading volume, and market capitalization. To enable statistical inference, we account for market-wide trends by considering counterfactual assets with similar historical patterns.

	\item We apply this method to our dataset and find a statistically significant negative price impact for \SigPriceImpactNegNPct/\% ($N=\SigPriceImpactNegN/$) governance assets as a result of the crime event. The affected DAOs experience an average price decline of \SigPriceImpactNegPct/\%. Also, for \SigVolImpactPosNPct/\% ($N=\SigVolImpactPosN/$), the trading volume of these assets increased significantly as a consequence of the event.

	\item Overall, the negative price effects of these DeFi crime events are reflected in the market capitalization of affected DAOs. We estimate that the indirect economic impact of DeFi crimes on targeted DAOs is at least \IndirectEcoImpact/ USD, far exceeding the direct costs experienced by the victims. Notably, the indirect economic impact accounts for \IndirectEcoImpactPct/\% of total losses.

\end{enumerate}

Hence, DeFi crime events have a significant impact on DAOs. When an effect is observed, the average price decline of approximately \num[round-mode=places, round-precision=0]{\fpeval{abs(\SigPriceImpactNegPct/)}}\% is substantially larger than the single-digit losses typically seen in traditional corporations~\cite{Woods2021, Ebrahimi2022}. Moreover, the indirect costs of these crimes far exceed the direct financial losses suffered by victims. These findings provide valuable insights for DAO investors and regulators, underscoring the importance of strengthening DeFi security given the magnitude of the economic harm.

Our study employs a rigorous, reproducible methodology applicable to DeFi crime events and governance tokens, surpassing industry reports and non-scientific analyses. By openly sharing our data and analytics framework at [\textit{{link accessible upon publication}}], we ensure transparency and encourage further research. Beyond assessing direct economic impacts, our findings provide key insights into how DAOs, as the novel organizational form for decentralized applications, respond to crime events, enhancing the understanding of market dynamics in decentralized governance.

The rest of the paper is organized as follows: Section~\ref{sec:Literature} reviews the literature on the impact of cybercrime on traditional organizations and the cryptoasset ecosystem. Section~\ref{sec:data} describes the data, and Section~\ref{sec:Method} outlines the methods. Section~\ref{sec:results} presents the study's results, while Section~\ref{sec:discussion} discusses the findings and highlights main limitations. Finally, Section~\ref{sec:conclusion} concludes the paper.


\section{The Economic Impact of Cybercrime}
\label{sec:Literature}

This section provides an overview of the current understanding of the impact of cybercrime on both traditional organizations and the cryptoasset ecosystem, which includes DeFi services. It highlights key findings from state-of-the-art research and introduces the fundamental concepts of Decentralized Autonomous Organizations (DAOs).

\subsection{On Traditional Organizations}

The economic impact of cybercrime extends beyond the direct financial gains of criminals or the immediate losses suffered by victims.
Previous research examined these indirect costs, including expenses related to strengthening cybersecurity measures and efforts to restore or maintain organizational trust~\cite{Anderson2013,Anderson2019}.
Additionally, another study demonstrated that profit-driven cybercrime affects organizations in multiple ways, with economic harm being a major consequence.
For example, market reactions may arise from cyber-related disruptions such as operational failures or deteriorating customer relationships.
Similarly, stock price declines may result from negative media coverage following the public disclosure of a cyber incident~\cite{Agrafiotis2018}.

In the economics literature, the impact of cybercrime on stock prices is explained through the efficiency of capital markets, which assumes that all available information is incorporated into prices via abnormal returns~\cite{Fama1970}. A systematic review of 45 event studies on information security found that the majority (75.6\%) reported a statistically significant impact on stock prices~\cite{Spanos2016}. Similarly, in an early study on stock market reactions, researchers examined short-term stock market responses and found that the effects were negative, though their magnitudes varied depending on the type of event~\cite{Campbell2003}.
Likewise, recent research analyzed financial market reactions to corporate security breaches in both the short and long term~\cite{Tosun2021}.
They observed declines in daily returns and increases in trading volume among affected corporations compared to similar, unaffected firms.
Although these breaches did not significantly impact long-term stock performance (up to five years), they did influence corporate policies, especially in areas such as payouts and R\&D.

In terms of specific figures, a literature review on the short-term effects of cybercrime events found that, out of 80 studies, 75\% reported a significant impact on companies' stock market performance~\cite{Ali2021a}. On average, stock prices declined by 3.5\%, with variations ranging from -0.25\% to -10\%, primarily within the two days before and after the event. In another study, researchers examined cyber risk from a broader perspective and conducted a meta-analysis of 19 studies on stock market reactions. The vast majority found negative short-term responses to cyber incidents and the cumulative abnormal returns (CAR) reported ranged from -0.8\% to -2.3\% within a one-day event window before and after the incident~\cite{Woods2021}. Similarly, a meta-analysis of 63 primary studies on security breach announcements reported an average CAR of -0.953\%~\cite{Ebrahimi2022}.

Event studies on market reactions typically conduct short-term analyses due to the difficulty of controlling for other influencing factors. However, research has examined the long-term effects of security breach disclosures. Using a one-to-one matching approach to compare affected and unaffected organizations, researchers found that affected organizations experienced negative abnormal returns between -15\% and -18\% within twelve months after the public disclosure of the breach~\cite{Ali2021}. Thus, while the reported price declines are substantial, their long-term nature introduces uncertainty due to other potential influencing factors.

In summary, the economic impact of cybercrime on traditional organizations has been extensively studied, with research consistently reporting price declines in the low single-digit percentage range.

\subsection{On Cryptoasset Markets}

An early study investigated the market impact of cryptocurrency thefts across 10 distinct assets. The authors found that between 2014 and 2019, short-term prices often increased following major theft incidents, attributing this phenomenon to heightened visibility driven by theft-related news coverage~\cite{Brown2020}.

In contrast, a study covering a similar period (2012–2021) found the opposite effect. Cyberattacks on cryptocurrency exchanges led to a statistically significant decline of 1.51\% in Bitcoin’s price on the day of the incident~\cite{Milunovich2022a}. However, this effect diminished in significance after 2019. Similarly, researchers demonstrated that hacking incidents negatively affected Bitcoin spot price returns, while also increasing volatility and reducing the price of future option products~\cite{Chen2023}. Additionally, a study analyzed 17 DDoS attacks on the centralized exchange Bitfinex, reporting that Bitcoin traded on the exchange exhibited significant negative abnormal trading volume in only four of these incidents~\cite{Abhishta2019}.

Conversely, cryptocurrency prices have shown positive market reactions to various external factors, including regulatory developments~\cite{Auer2018}, law enforcement actions~\citep{Abramova2021}, and social media activity by prominent figures~\cite{Ante2023}. Moreover, cryptocurrency-specific events, such as major exchange listings or delistings (i.e., the removal of a cryptocurrency from an exchange, making it unavailable for trading), as well as airdrop announcements (i.e., the distribution of assets to individual addresses), have been shown to trigger significant abnormal returns. Enforcement actions by the U.S. Securities and Exchange Commission (SEC) have also played a pivotal role in influencing market movements~\cite{Auer2018}.

Notably, negative events tend to exert a greater impact on cryptocurrency prices than positive ones~\citep{Emrah2022}. For instance, a study examined the effects of 173 cybersecurity breaches on crypto token projects using market data from the aggregator CoinGecko between 2020 and 2023. The findings revealed significant negative abnormal returns, with the most pronounced declines observed in financial and smaller projects with lower market capitalization~\cite{Li2024}.

In summary, the literature on cybercrime events' impact on cryptoasset markets is more limited and less consistent than that in traditional finance, with findings sometimes presenting contradictory effects, such as price increases in some cases and decreases in other cases.

\subsection{Decentralized Autonomous Organizations (DAOs)}

Decentralized Autonomous Organizations (DAOs) represent a novel organizational model for applications operating in decentralized environments, functioning without central control or management~\cite{Liu2021,Kiayias2023}. Vitalik Buterin initially conceptualized DAOs as a framework for the long-term management of smart contracts governing digital assets~\cite{Buterin2014}.

In Ethereum-like distributed ledgers, cryptoassets are implemented as executable software programs known as smart contracts.
Also referred to as tokens, cryptoassets represent real-world assets or rights on the blockchain.
This innovation gained momentum with Decentralized Finance (DeFi), which enables financial applications for exchanging or lending tokens in a decentralized manner~\cite{Auer2024,Harvey2020}.
Among the most widely used DeFi services are decentralized exchanges (DEXs), which facilitate the automated exchange of tokens without an intermediary~\cite{Xu2022a,Angeris2024}.
Internally, DeFi services are typically structured through multiple smart contracts, including treasury contracts that hold tokens worth billions of dollars in Total Value Locked (TVL)~\cite{Saggese2025}.
The combination of substantial reserve values and the increasing complexity of DeFi underscores the need for robust governance mechanisms.

DAOs provide an organizational framework for these services, enabling decentralized decision-making through community voting on improvement proposals. To facilitate this process, DAOs issue dedicated governance tokens that grant voting rights, with decision-making power proportional to the number of tokens a voter holds~\cite{Laturnus2023,Barbereau2022}.

However, recent literature questions the decentralized nature of DAOs~\cite{Tan2023, Kitzler2023a}, highlighting issues such as power concentration and unexercised voting rights~\cite{Dotan2023a, Fritsch2022a}. Additionally, these relatively new type of organization faces various challenges~\cite{Tan2023} and governance attacks~\cite{Feichtinger2024}, including exploits targeting vulnerabilities. Although DAOs issue and distribute governance tokens, these tokens can be traded on exchanges, making them functionally similar to shares in traditional organizations.

Despite these concerns, only a single study has examined the effect of decentralized governance mechanisms on their underlying applications. Specifically, this research identified a relationship between governance and performance, showing that governance measures aimed at enhancing security are associated with positive abnormal returns~\cite{Appel2023a}.

In summary, existing studies on DAOs primarily focus on analyzing their decentralized nature and examining governance attacks.

\subsection{The Present Study}

Previous research has documented the economic impact of cybercrime events on both traditional organizations and cryptocurrencies. While market reactions typically lead to price declines for companies, the effect on cryptocurrencies remains less clear and inconsistent.

However, the impact of cybercrime on the novel organizational form of DAOs, as reflected in the market reactions of their governance tokens, remains largely unexplored. Existing event studies on cryptoassets primarily focus on abnormal asset returns, but often overlook counterfactual trends of similar assets, limiting causal inference. Additionally, most studies rely on off-chain market price data from centralized exchanges, which lack granularity and do not capture settled transactions.
Our study addresses these gaps.


\section{Data}
\label{sec:data}

To assess how DeFi crime events impact prices, trading volumes, and market capitalization of governance tokens, requires multiple data sources. We present the dataset preprocessing steps below.

\subsection{DeFi Crime Events}

To select crime events, we used a curated database on cybercrime in the DeFi space, which includes data on \num{1141} crime events from 2017 to 2022~\cite{CarpentierDesjardins2025}. Of these, \num{1036} events are related to DeFi services. The database aggregates information from news platforms such as Slowmist and DeFi.Rekt. For each crime event, it includes: i) the DeFi actor involved (e.\,g., DEX or lending service), ii) the direct USD financial damage resulting from the attack, and iii) an estimate of the date when the attack was announced publicly. Furthermore, the dataset is available online~\cite{carpentier-desjardins2024mapping} and provides a categorization of the strategies and tactics used in each crime event.

\subsection{Governance Tokens}

To identify DAO-governed DeFi actors in this dataset, we retrieved a list of all known governance tokens using the CoinGecko API\footnote{https://www.coingecko.com/en/api, accessed on 2024-01-29}. Since the majority of DeFi activity has been concentrated on the Ethereum network (accounting for over 60\% of Total Value Locked (TVL) until 2022\footnote{\url{https://defillama.com/chains}}), we restricted our selection to tokens on this network. For each token, we collected its Ethereum address, name, and symbol. This resulted in a list of \CoinGeckoGovTokensN/ known governance tokens.

We then merged this list with the DeFi actors in the crime event dataset. This process was carried out through an automated matching procedure, which compared governance token names with DeFi actor names, followed by manual verification (see \ref{par:matching}). After this filtering step, we generated a preliminary list of \GovOverlapsN/ governance tokens and \DAOEventsN/ associated DeFi crime events.

In the following, we denote each crime event as $e = (a,\tau)$, where $a$ represents the governance asset of the compromised DAO, and $\tau$ denotes the date of the event's public announcement.

\subsection{Trading Volumes and Prices}

We obtained on-chain trading data from Uniswap V2, the most prominent Decentralized Exchange (DEX) operating across multiple blockchain networks, which was the leading platform during the analyzed event period.\footnote{\url{https://defillama.com/trending-contracts}}. It uses an Automated Market Maker (AMM) model based on paired-asset liquidity pools (LPs) for trading and price formation~\cite{Xu2022a,heimbach2021behavior}. As before, we restricted our analysis to Ethereum and ran a full archive node to retrieve contract information and log data.

In the following, we describe our process for identifying relevant liquidity pools, extracting trading volume and price data, and performing necessary steps to compile the final dataset.

\subsubsection{Identifying Liquidity Pools}

First, we retrieved all deployed liquidity pool addresses
of the Uniswap V2 factory contract\footnote{\textit{0x5c69bee701ef814a2b6a3edd4b1652cb9cc5aa6f} created on block 10000835 (2020-05-04)},
using the Web3 interface of the Ethereum node. The factory contract is responsible for creating and managing LPs.
At the time of writing, we retrieved \LPoolsN/ distinct liquidity pools. To compute trading volumes and prices, we restricted our selection to LPs of Wrapped Ether (wETH) and governance token $a$ pair combinations $(wETH, a)$. The rationale behind this selection is that such pools exist for most governance tokens and provide high liquidity and frequent trading activity.

The analyses are conducted in the native currency, but exchange rates (e.g., ETH-USD) are generally available, allowing for easy conversion to fiat values. In total, we identified \LPoolsGovN/ liquidity pools of type $(wETH, a)$ for governance tokens.

\subsubsection{Extracting Volumes and Prices}

Next, we extracted trading volumes and prices of governance assets from the logs of the Ethereum full archive node. We retrieved raw logs for each liquidity pool, starting from the deployment of Uniswap V2, covering blocks \num{10000835} (2020-05-04) to \num{16301000} (2022-12-30). We then filtered for \textit{swap} and \textit{sync} event logs. The swap event is emitted when the \textit{swap} function of a Uniswap LP contract is called and a trade is settled. It provides the traded volume $v_{a}$ of the governance asset in $wETH$. The \textit{sync} event is emitted whenever liquidity volumes $(L_{wEth},L_{a})$ in the pool change.

To compute prices, we followed the method proposed by \cite{Adams2020}, using pool liquidity volumes from \textit{sync} events, such that: $p_{a} = \frac{L_{a}}{L_{wETH}}$, where $p_{a}$ is the price of the governance asset in the native currency $wETH$. Since these function calls are executed within the scope of a specific Ethereum transaction, which is embedded in a block, we used the block timestamp. Consequently, for each governance asset $a$ and timestamp $t$, we obtained the data points price $p_{a,t}$ and volume $v_{a,t}$. Note that multiple trades can occur within a single block, potentially resulting in several volume and price data points for the same timestamp.

\subsubsection{Data Pre-processing}

Our procedure for extracting volumes and prices generates data points at the granularity of individual trades, with potentially irregular timing patterns including periods of sporadic transactions or concentrated bursts of trading activity.
However, our event study methodology requires discrete prices or volumes aggregated into specific time intervals. Additionally, the data may include non-representative trading patterns, such as flash loans or large arbitrage trades, which can distort prices due to a few deviating transactions~\cite{Lehar2022}. To mitigate this, we identified and removed such outliers before aggregating the extracted trading data.

To detect large deviations, we applied the base-10 logarithm to the price series and compared individual values to the rolling median (centered over \num{1000} trades) using the modified Z-score, which is a statistical measure based on the median rather than the average. We began with the entire series and flagged prices with a modified Z-score above \num{3.5}, a commonly used threshold for identifying outliers~\cite{Iglewicz1993}.

Next, we assessed the local price environment to ensure that detected deviations were isolated trades rather than part of broader trends. Specifically, we examined five blocks before and after each deviation, removing values exhibiting spiking behavior, which is characterized by abrupt increases or decreases that quickly reverted. This filtering was performed again using the modified Z-score, but calculated within a local window of \num{100} preceding and succeeding trades.

\subsubsection{Data Aggregation}

To construct the series, we aggregated governance token prices and volumes by first defining an interval $\Delta t$ and accumulating all trades executed within the time window $[t,t+\Delta)$.

For prices, we computed the median price within each interval, as the median is more robust than the mean against fluctuations induced by trading activity. Additionally, we handled missing time slots using the last observation carried forward (LOCF) method. For volumes, we summed the total traded volume within each interval, irrespective of whether the trades were buys or sells.

Crime announcement dates $\tau$ lack specific time-of-day and timezone information. Therefore, we assume midnight UTC but account for potential timing deviations of the market reaction by analyzing observations over a time window around the announcement date. This approach allows us to study the temporal dynamics rather than relying on a single point-in-time observation.

For each crime event $e$, we verified the availability of price and volume data both on the date of public announcement and during the preceding reference window.
Additionally, we removed sparse time series, defined as cases where less than one annual cycle of data was available or the average trading frequency was below one trade per day (see \ref{ssec:Clearning}). In total, prices and volumes of \GovPricesN/ governance tokens were extracted.

\subsection{Market Capitalization}\label{ssec:MarketCap}

We calculated the on-chain market capitalization of a governance token $a$ at timestamp $t$ by multiplying its token supply $s$ with the price $p_{a,t}$. To obtain the supply, we queried the Web3 interface of our local Ethereum node, calling the \textit{totalSupply()} function at the corresponding block height. We then converted the price from ETH to USD using the exchange rate from Uniswap LP $(wETH, USDC)$, assuming a constant USD peg of the USDC stablecoin.

\subsection{Dataset Summary}

\begin{figure}[!htb]
	\centering
	\includegraphics[width=1.\columnwidth]{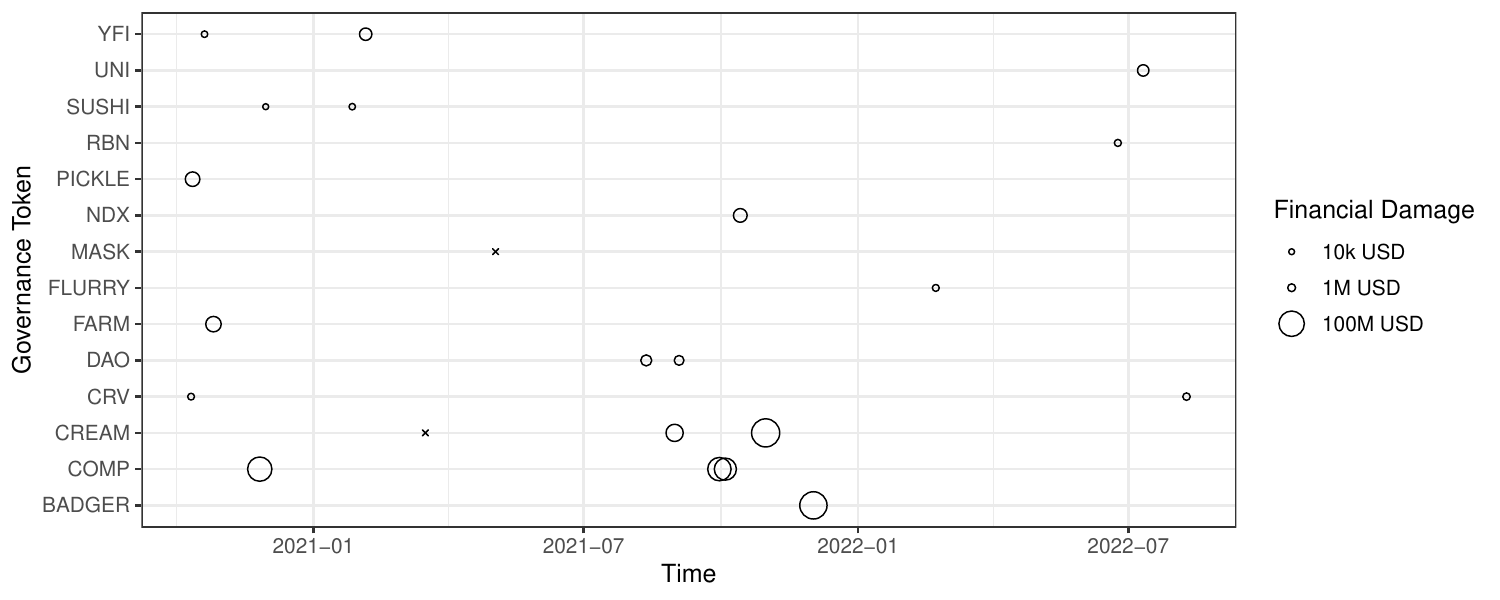}
	\caption{\textbf{DAO Crime Timeline}: This figure visualizes the  announcement dates of DeFi crime events and their affected governance tokens. The financial damage (direct economic impact) is represented by bubble size, or a cross-mark if the information is unavailable. }
	\label{fig:plEventTimeLine}
\end{figure}

In total, we curated a price and volume dataset of \GovPricesN/ governance tokens to conduct inference analysis with controls. From this dataset, we identified \UniqueGovAssetsN/ governance tokens linked to \DAOEventsAvailableN/ crime events. Figure~\ref{fig:plEventTimeLine} provides a temporal overview of governance assets, associated crime events, and their direct economic impact (referred to as direct financial impact by \cite{carpentier-desjardins2024mapping}).

Table~\ref{tab:crimeEvents} provides further details on the crime events, including the associated governance assets, their direct economic impact, and the targeted DAOs. Approximately 50\% (N=11) of these incidents involved attacks exploiting contract vulnerabilities, often due to flawed access control mechanisms or logical bugs in the smart contract code. These incidents alone resulted in direct financial damages exceeding \$367 million. Other types of attacks included DNS attacks (N=3), phishing campaigns (N=2) that compromised infrastructures, flash loan exploits (N=3), oracle manipulations via price feeds (N=1), a compromised API key (N=1), and an attack on a token vault (N=1). Further details on attack strategies and specific tactics can be found in \cite{CarpentierDesjardins2025}.
Overall, according to the authors of the original dataset, these \DAOEventsAvailableN/ crime events targeting DAOs resulted in \$613 million in direct financial damages~\cite{CarpentierDesjardins2025}.

\renewcommand{\arraystretch}{1.}

\begin{table}[!htb]
	\centering
	{
	\small
	\begin{tabular*}{\textwidth}{@{\extracolsep{\fill}}
		p{0.25cm}
		p{1.5cm}
		p{1.25cm}
		m{6.cm}
		>{\centering\arraybackslash} p{1.7cm}
		>{\raggedleft\arraybackslash} p{1.8cm} }
\toprule
\textbf{\#} & \textbf{Gover- nance Asset $a$} & \textbf{DAO} & \textbf{Crime Event $e$} & \textbf{Announce- ment $\tau$} & \textbf{Direct Economic Impact}\\
\midrule
1 & CRV & Curve & Phishing attack on the exchange website  & 2020-10-11 & 227\,k\\
2 & PICKLE & Pickle & \strut Attack through contract vulnerability, access control flaw \strut & 2020-10-12 & 19.7M\\
3 & YFI & Yearn &  \strut Phishing scam imitating YFI \strut & 2020-10-20 & 129.5k\\
4 & FARM & Farm & \strut Attack through contract vulnerability, logical bug/custom flaw \strut & 2020-10-26 & 24M\\
5 & COMP & Compound &  Oracle attack through price feed & 2020-11-26 & 89M\\
6 & SUSHI & Sushi & Attack through contract vulnerability, logical bug/custom flaw & 2020-11-30 & 15k\\
7 & SUSHI & Sushi & Attack through contract vulnerability, logical bug/custom flaw & 2021-01-27 & 105.2k\\
8 & YFI & Yearn & Attack on yDAI vault & 2021-02-05 & 11M\\
9 & CREAM & Cream & DNS attack  & 2021-03-17 & -\\
10 & MASK & Mask & Attack through contract vulnerability & 2021-05-03 & -\\
11 & DAO & DAO Maker & Attack through contract vulnerability, access control flaw & 2021-08-12 & 7M\\
12 & CREAM & Cream & Attack through contract vulnerability, reentrancy & 2021-08-31 & 34M\\
13 & DAO & DAO Maker & Attack through contract vulnerability, access control flaw & 2021-09-03 & 4M\\
14 & COMP & Compound & Attack through contract vulnerability, logical bug/custom flaw & 2021-09-30 & 80M\\
15 & COMP & Compound &  Attack through contract vulnerability, access control flaw & 2021-10-04 & 68.8M\\
16 & NDX & Indexed & Attack via flash loans  & 2021-10-14 & 16M\\
17 & CREAM & Cream &  Attack through contract vulnerability, logical bug/custom flaw  & 2021-10-31 & 130M\\
18 & BADGER & Badger & Compromised API key enabled script injection on the protocol's front end & 2021-12-02 & 120M\\
19 & FLURRY & Flurry & Attack via flash loans & 2022-02-22 & 293k\\
20 & RBN & Ribbon &  DNS attack & 2022-06-24 & 351.1k\\
21 & UNI & Uniswap & Airdrop phishing scam & 2022-07-11 & 8M\\
22 & CRV & Curve &  DNS attack & 2022-08-09 & 612k\\
\midrule
& & & & & 613.2M \\
\bottomrule
\end{tabular*}

	}
	\caption{\textbf{Crime Events and Governance Assets}:
		This table lists crime events including the associated governance asset $a$, DAOs, and a brief summary of the crime event $e$. Events are ordered by their announcement date $\tau$, with the direct economic impact reported in US dollars (if available). }
	\label{tab:crimeEvents}

\end{table}


\section{Methods}
\label{sec:Method}

To quantify the indirect economic impact of DeFi crime events on DAOs, we employ an event study framework. Specifically, we use a dynamic difference-in-differences (DiD) model to analyze the effects of these events on both prices and trading volumes within a short temporal window surrounding each crime event. Unlike traditional approaches, such as the abnormal return model~\cite{Brown1985,Campbell1996}, our method explicitly incorporates broader market dynamics. To strengthen causal inference, we account for market-wide trends by incorporating counterfactual assets with similar historical behavior. For clarity, we proceed as follows: we first define the three time windows created for the event analyses. We then describe the method used to identify counterfactual assets similar to those affected by crime events. Finally, we introduce the core concepts of the DiD model, incorporating the chosen time windows and identified counterfactual assets.

\subsection{Time Windows}
\label{subsec:time_windows}

The event study framework conceptually analyzes the impact of an event by comparing an asset's price and volume during a specified \emph{event window} $W^{\text{E}}$ with values from a \emph{reference window} preceding the event. For the reference window, we distinguish between two periods:

\begin{enumerate}
	\item The long window $W^{\text{L}}$ is used to identify counterfactuals with similar market behavior.
	\item The short window $W^{\text{S}}$ captures the immediate period preceding the announcement of a DeFi crime event.
\end{enumerate}

\begin{figure}[!t]
	\includegraphics[width=1\textwidth]{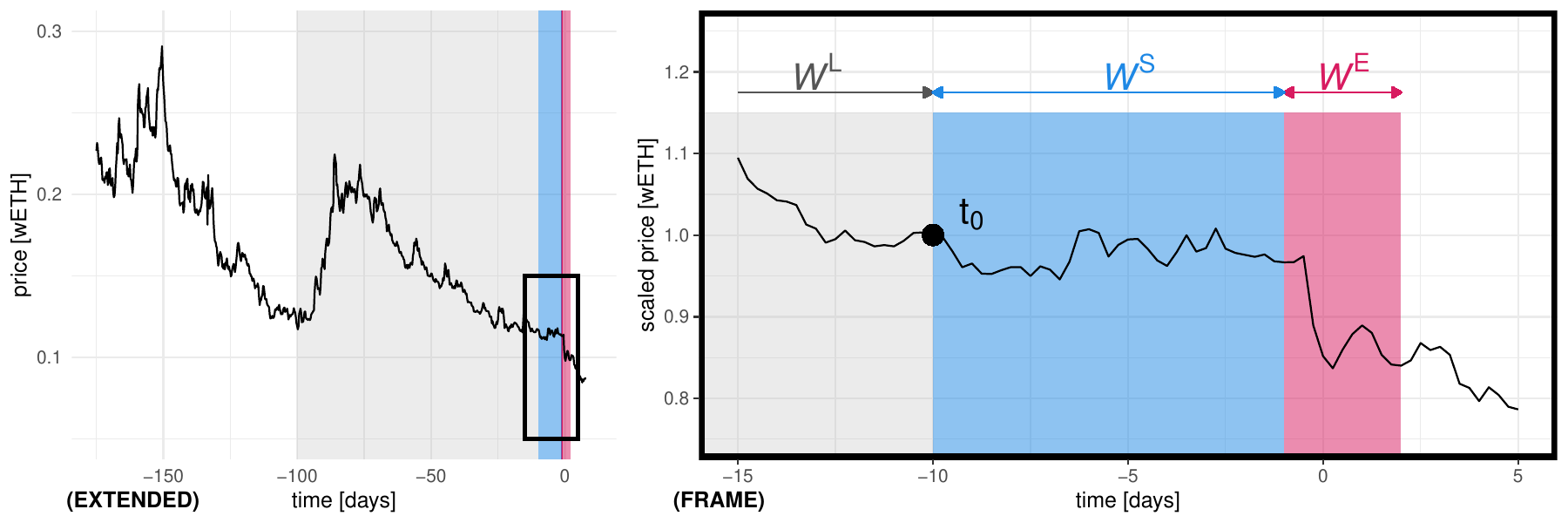}
	\caption{\textbf{Event Study Framework}: The time series is divided into three windows: the event window $W^{\text{E}}$, which surrounds the crime event, and two reference windows. The long reference window $W^{\text{L}}$ is used to identify counterfactual assets, while the short reference window $W^{\text{S}}$ is used to normalize the series at the beginning, assuming a parallel trend before the event. The right plot provides a zoomed-in view of the framed section from the time series shown in the left panel.}
	\label{fig:pl_event_concept}
\end{figure}

Figure~\ref{fig:pl_event_concept} illustrates the conceptual framework. It depicts a time frame (right) overlaid on the time series (left), highlighting the event window along with the short and long reference windows. The specific durations of $W^{\text{E}}$, $W^{\text{L}}$, and $W^{\text{S}}$, measured in days, are detailed in the parameterization section below.

\subsection{Identifying Counterfactual Assets}
\label{ssec:counterassets}

A DiD model compares assets that exhibit parallel trends in a representative target variable $Y$, which, in our case, includes price $p$ and trading volume $v$. When an intervention occurs on an asset $a$ --- for instance, a crime event $e$ --- the effect can be analyzed relative to other assets, referred to as counterfactual assets $\tilde{A_e}$. These assets initially display similar market trends and are presumed unaffected by the event of interest.

\subsubsection{Selection}

To identify counterfactual assets while accounting for the diversity of DAOs, we restrict our analysis to the same asset class, i.e., governance tokens, and further refine the selection using a three-step approach.

First, we standardize the variables $Y_{a,t}$ to $Z_{a,t}$ using the z-score metric, as defined in Equation~\ref{eq:Zscore}. This procedure is applied to the long reference window $W^{\text{L}}$ preceding the event, which is also used to compute the mean $\mu$ and standard deviation $\sigma$ of the series within this window. Standardization ensures that we capture assets with initially similar behavior, even if their amplitudes differ.

\begin{equation}\label{eq:Zscore}
	Z_{a,t} := \text{Zscore}(Y_{a,t}) = \frac{Y_{a,t} - \mu}{\sigma}
\end{equation}

Second, we compute the Pearson correlation $r_{a_e,a_x}$ between the targeted governance asset $a_e$ (associated with the crime event) and all other governance assets $a_x \in A \setminus \{a_e\}$. To do so, we correlate values $Z_{a_e,t}$ and $Z_{a_x,t}$ for the available pairs at $t \in W^{\text{L}}$. We then construct an interim ordered list of assets $\tilde{A_e}^{(1)}$, consisting of those that exceed a minimum correlation threshold $t_r$.

\begin{equation}
	\tilde{A_e}^{(1)} = \{a_x \, \mid \, (r_{a_e,a_x} > t_r) \} \quad.
	\label{eq:counterf1}
\end{equation}

Third, we limit the number of counterfactual assets to at most $c$ per DeFi crime event and targeted governance token. Including too many control assets may introduce noise rather than capturing the true effect. To refine the selection, we rank the assets in $\tilde{A_e}^{(1)}$ by correlation and retain only the top $c$ assets (Equation~\ref{eq:top_cf}). To further eliminate less correlated assets, we compute the mean correlation of the selected top $c$ assets and remove those falling below this threshold (Equation~\ref{eq:top_cf_refined}).

\begin{equation}
	\tilde{A_e}^{(2)} = \{a_x' \mid a_x' \in \tilde{A_e}^{(1)}, x \in \{1, \dots, n\}, n \leq c \}
	\label{eq:top_cf}
\end{equation}

\begin{equation}
	\begin{aligned}
		\tilde{A_e} &:= \tilde{A_e}^{(3)} = \{a_x'' \mid r_{a_e,a_x''} \geq \text{mean}(R^{(c)}) \},  \\
		&\text{where} \, R^{(c)} = \{r_{a_e,a_x'} \mid \ a_x' \in \tilde{A_e}^{(2)}\}.
	\end{aligned}
	\label{eq:top_cf_refined}
\end{equation}

\subsubsection{Scaling}

Finally, as a preparation step for the DiD analysis, we take the set of identified counterfactual assets and normalize their prices and volumes to $\hat{Y}_{a,t}$. Scaling ensures that the (price, volume) time series in the event window are comparable to the baseline in the reference window.

To achieve this, we normalize at the beginning of the short estimation period $W^{\text{S}}$ and scale each series relative to its prior value $t_{s}$ (see dot in Figure~\ref{fig:pl_event_concept}). The identified counterfactual assets are then used in the DiD analysis to compare the effect of the crime event on a governance asset against other, similar assets.

\subsection{Dynamic Difference-in-Differences (DiD) Model}

The dynamic difference-in-differences (DiD) model developed in this study compares the target variable $\hat{Y}_{a,t}$ (i.e., price or trading volume) of a specific governance asset $a_e$, which was affected by a crime event $e$, against the set of identified counterfactual assets $\tilde{A_e}$.

As introduced in the standard literature~\cite{HuntingtonKlein} and applied in empirical economics research~\cite{Ershov2020EC}, we adapt the common DiD design for dynamic effects, as defined in regression Equation~\ref{eq:dDidre}. For each crime event, we model the normalized dependent variable $\hat{Y}_{a,t}$ for governance and counterfactual assets $a \in (\{a_e\} \cup \tilde{A_e})$ at time $t$, relative to the announcement date $\tau_e$.

\begin{equation}\label{eq:dDidre}
	\hat{Y}_{a,t}  = \, \alpha_0 + \beta_a + \beta_t + \sum_{t' \in T^E \setminus t_a} \gamma_{t'} \, D_{a,t'} + \epsilon_{a,t}
\end{equation}

We use fixed effects for assets $\beta_a$ and time $\beta_t$ and evaluate the dynamic effect of $e$ using the regression coefficients $\gamma_{t'}$ over time $t' \in T^E \setminus \{t_a\}$. We exclude the anchor time $t_a$ before the announcement date $\tau_e$, as it serves as the reference point to which the effect ($\gamma_{t'}$) is compared. For the $\gamma_{t'}$ coefficients, the regression includes an interaction term $D_{a,t'}$, constructed from time and asset as follows:

\begin{equation}
	D_{a,t'} := \begin{cases}
		1 & \text{if} \, (a = a_e) \, \text{and}\, (t' = t) \\
		0 & \text{otherwise}
	\end{cases}
	\quad.
\end{equation}

Moreover, we cluster standard errors at the asset level $a$ to account for within-entity correlation, i.e., time series autocorrelation, ensuring robust inference. Using the standard errors of the $\gamma_{t'}$ coefficients, we assess the statistical significance of the effects.

In summary, our DiD approach allows us to analyze the temporal dynamics of crime event effects, specifically when the effect occurs and whether it persists or fades over time, and get closer to causal interpretation.

\subsection{Model Parametrization}

Prior event studies on cryptocurrencies~\cite{Auer2018,Abramova2021} have used short event windows to account for high volatility and rapid market reactions. Consistently, we define the event window as $W^{E}  = [-1, +2) $ days, setting $t_a = -1$ day as the pre-event baseline before the effect occurs. For the reference window, we distinguish between long and short periods:

\begin{enumerate}

	\item The long window $W^{L}$ spans $[-100, -10)$ days and is used to identify counterfactuals by analyzing market behavior over an extended period.

	\item The short window $W^{S}$ encompassing $[-10, -1)$ days and captures the immediate period preceding the announcement of a DeFi crime event.

\end{enumerate}

Finally, we set the aggregation interval to $\Delta t = 6$ hours to capture intraday market reactions and select $p < 0.1$ as the threshold for statistically significant effects in the DiD analysis. For the counterfactual asset identification procedure, we limit the number of counterfactual assets to $c = 10$ ($\approx10\%$ total of \GovPricesN/ governance assets). Additionally, we set a minimum correlation threshold of $t_r = 0.4$. \ref{ssec:AppendixCounterfactuals} presents results for different thresholds and demonstrates that our selected threshold maximizes the mean correlation while still ensuring at least one counterfactual asset per crime event.


\section{Results}
\label{sec:results}

In the following, we present the identified counterfactual assets and analyze the impact of DeFi crime events on prices, trading volumes, and the market capitalization of governance tokens.

\subsection{Identified Counterfactual Assets}

\begin{figure}[!t]
	\centering
	\includegraphics[width=1\columnwidth]{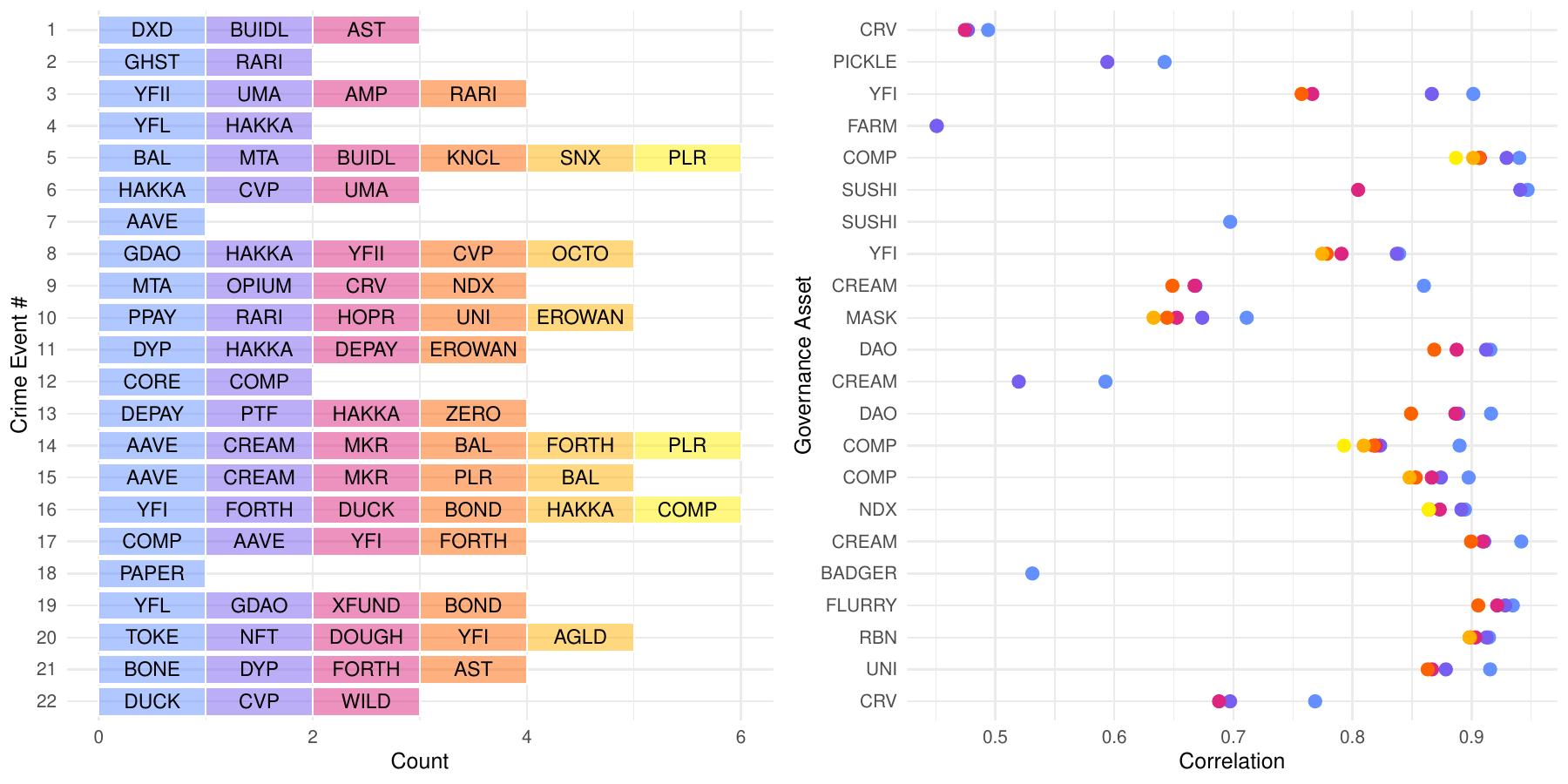}
	\caption{\textbf{Governance and Counterfactual Assets}: For each crime event $e$, the identified counterfactual assets (left) correspond to the targeted governance asset $a$, based on their historical price correlation (right) during the reference period $W^{L}$.}
	\label{fig:cor}
\end{figure}

Applying the method described in Section~\ref{ssec:counterassets} separately to prices and volumes for all DAO crime events listed in Table~\ref{tab:crimeEvents} resulted in a variable number of counterfactual assets. Figure~\ref{fig:cor} presents the counterfactual assets identified based on price dynamics, while \ref{ssec:AppendixCounterfactuals} provides the results for volumes. The analysis reveals that, for prices, we identified between one and six counterfactual assets per event. Notably, for many events, the procedure produced counterfactual assets with correlations significantly higher than the minimum threshold. On average, we identified $\PriceCorAvgN/ \pm \PriceCorSEN/$ counterfactual assets with correlation of $\PriceCorAvg/ \pm \PriceCorSE/$ for prices; and $\VolCorAvgN/ \pm \VolCorSEN/$ counterfactual assets with a correlation of $\VolCorAvg/ \pm \VolCorSE/$ for volumes.

\begin{figure}[!t]
	\includegraphics[width=0.95\textwidth]{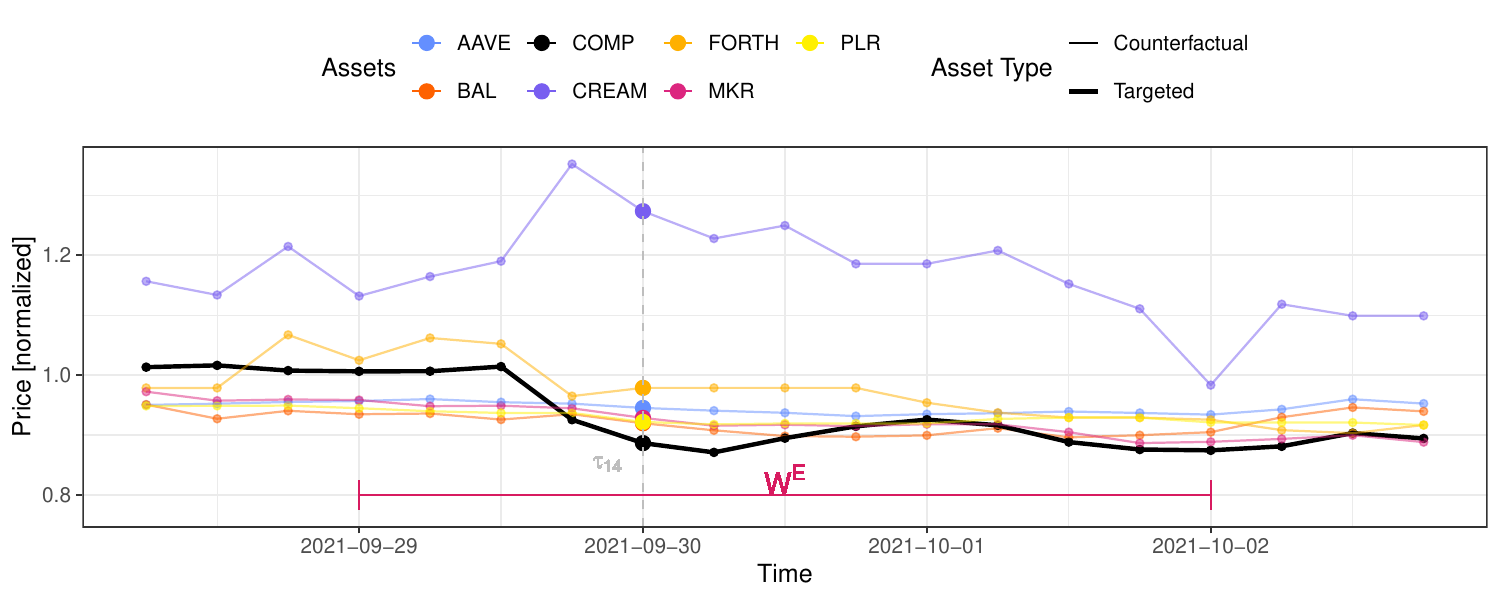}
	\caption{\textbf{Price Trajectory of \textit{Compound} Governance Asset and Related Counterfactual Assets}: This figure illustrates the price decline of the governance asset during crime event $e_{14}$ on \textit{Compound} DAO within the event window $W^E$. The identified counterfactual assets, selected based on the highest correlation in the reference period $W^{L}$, provide a benchmark to assess the economic impact of the crime.}
	\label{fig:plRegC1}
\end{figure}

As an example, Figure~\ref{fig:plRegC1} shows the price trajectory, $\hat{p}_{a_{14},t}$ for the governance asset $a_{14} = \textit{COMP}$ and its related counterfactual assets, given crime event $e_{14}$. Prior to the crime event, most assets in the plot exhibit a parallel trend. However, during the event window $W^E$, the governance asset $a_{14}$ experienced a price decline. By comparing the price trend of governance token $a_{14}$ with those of its counterfactual assets $\tilde{A_{14}}$, we infer the indirect economic impact of the crime.

\subsection{Impact of DeFi Crime Events on Prices}

To assess the impact of a DeFi crime event on the price of a targeted governance asset, we apply the DiD model to its price evolution, considering counterfactual assets.

\begin{figure}
	\centering
	\includegraphics[width=1\columnwidth]{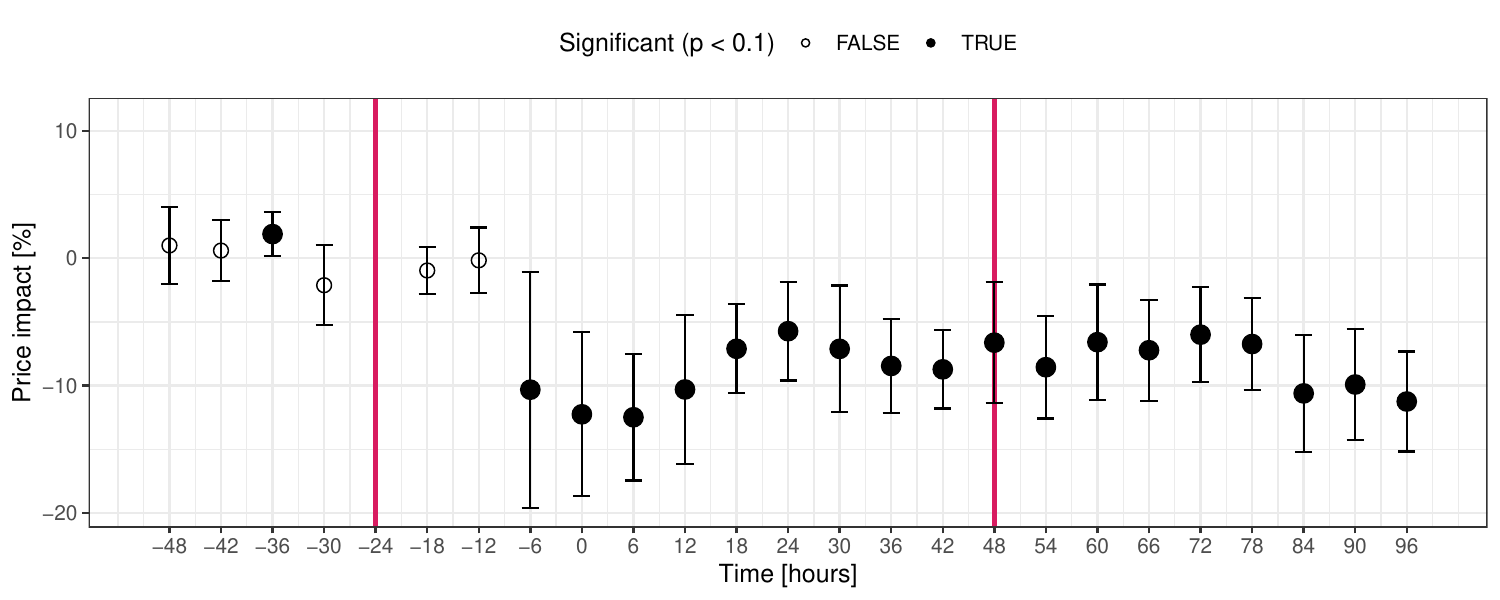}
	\caption{\textbf{Impact of DeFi Crime Event $e_{14}$ on the \textit{Compound} Governance Asset Price}: During the event window, the governance asset experienced a statistically significant price decline. The plot presents the estimated coefficients $\gamma_{t'}$ along with their $90\%$ confidence intervals, capturing the impact across the time frame surrounding $\tau_{14}$.}
	\label{fig:coefPlotC1}
\end{figure}

To facilitate comprehension, we first apply the method to a single DeFi crime event, the previously introduced \textit{Compound} DAO (crime event $e_{14}$). For this event, the regression coefficients $\gamma_{t'}$ computed using the DiD model, along with their $90\%$ confidence intervals, are presented in Figure~\ref{fig:coefPlotC1}. These coefficients measure the relative price impact of the crime event $e_{14}$ considering the asset's price at $t_a=-1$ day, the anchor time marking the beginning of the event window. Additionally, the two red lines indicate the event window $W^E$ used to calculate the economic impact. The first red line also marks the anchor time $t_a$, serving as the zero-reference point for the regression coefficients.

This example demonstrates that the price of the \textit{Compound} governance asset experienced a drop (compared to its counterfactual assets) around $t'=-6$h prior to the event announcement. This early decline could be attributed to shareholders gaining awareness of the event before its official announcement. It could also be a time deviation resulting from the conversion of date $\tau_{14}$ to daytime. This highlights the importance of selecting an appropriate time window when assessing the impact of a crime event. In this case, the price decline persists beyond the event window. Furthermore, relative to the anchor point $t_a$, the estimated price impact of the governance asset at $t'=-6$h is approximately -10\% ($\gamma_{t'=-6\text{h}} = \text{\EventExpMinusSixHoursEstimate/} \pm \text{\EventExpMinusSixHoursSE/}$).

\begin{figure}
	\centering
	\includegraphics[width=1.\textwidth]{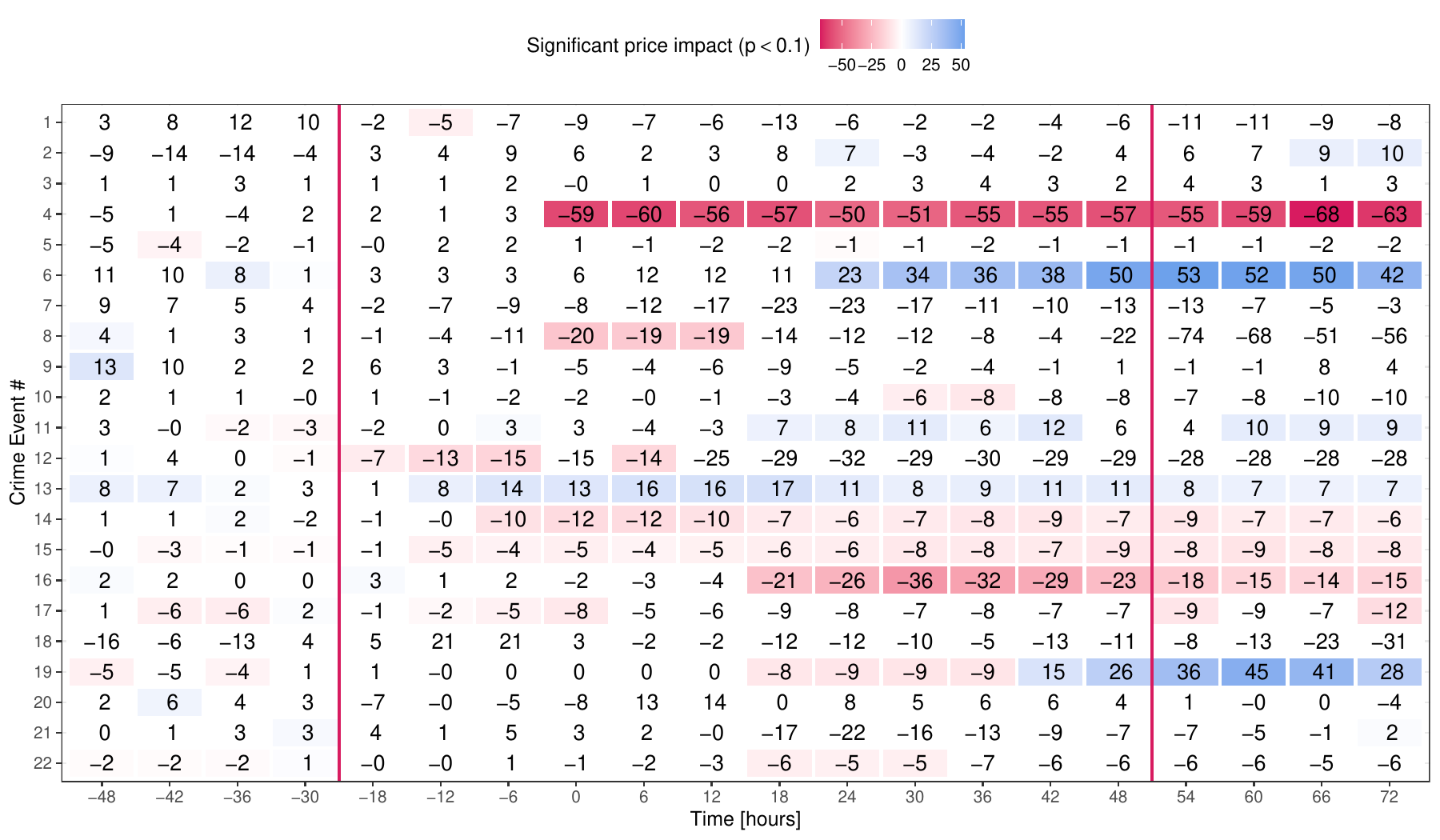}
	\caption{\textbf{Impact of DeFi Crime Events on DAO Governance Token Prices}: Each row represents the impact of a specific crime event on the associated DAO governance token price, with statistically significant values color-coded. \SigPriceImpactNegN/ events exhibit significantly negative price impact over the event window $W^{E} = [-1, +2]$ days.}
	\label{fig:plDynDiDheat}
\end{figure}

We then applied this method to all crime events listed in Table~\ref{tab:crimeEvents}. For simplicity, Figure~\ref{fig:plDynDiDheat} presents the estimated coefficients at different time points, with further details provided in \ref{ssec:AppendixDiD}. More specifically, Figure~\ref{fig:plDynDiDheat} illustrates the temporal price impact (i.e., the estimated coefficient $\gamma_t$, representing the percentage change in price) for each event $e$ (displayed row-wise). Notably, the dynamic DiD analysis, which evaluates price changes every six hours, offers granular insights into intraday variations and showing whether the impact persists over time. The two vertical red lines indicate the event window $W^E$, marking the area for potential occurring effects. Statistically significant coefficients are highlighted and color-coded.

\emph{
Overall, we found that \PriceTrendNeg/ out of the \DAOEventsAvailableN/ crime events exhibited a negative price trend over the event window. However, a negative and statistically significant effect ($p < 0.1$) on governance asset prices was observed in \SigPriceImpactNegNPct/\% (N = \SigPriceImpactNegN/) of the crime events, at least at one time step. For these events, the average statistically significant price decline was $\SigPriceImpactNegPct/\ \pm \SigPriceImpactNegSE/\%$, with a range from  \SigPriceImpactNegMin/\% to \SigPriceImpactNegMax/\%.
}

\subsection{Impact of DeFi Crime Events on Trading Volumes}

\begin{figure}
	\centering
	\includegraphics[width=1.\textwidth]{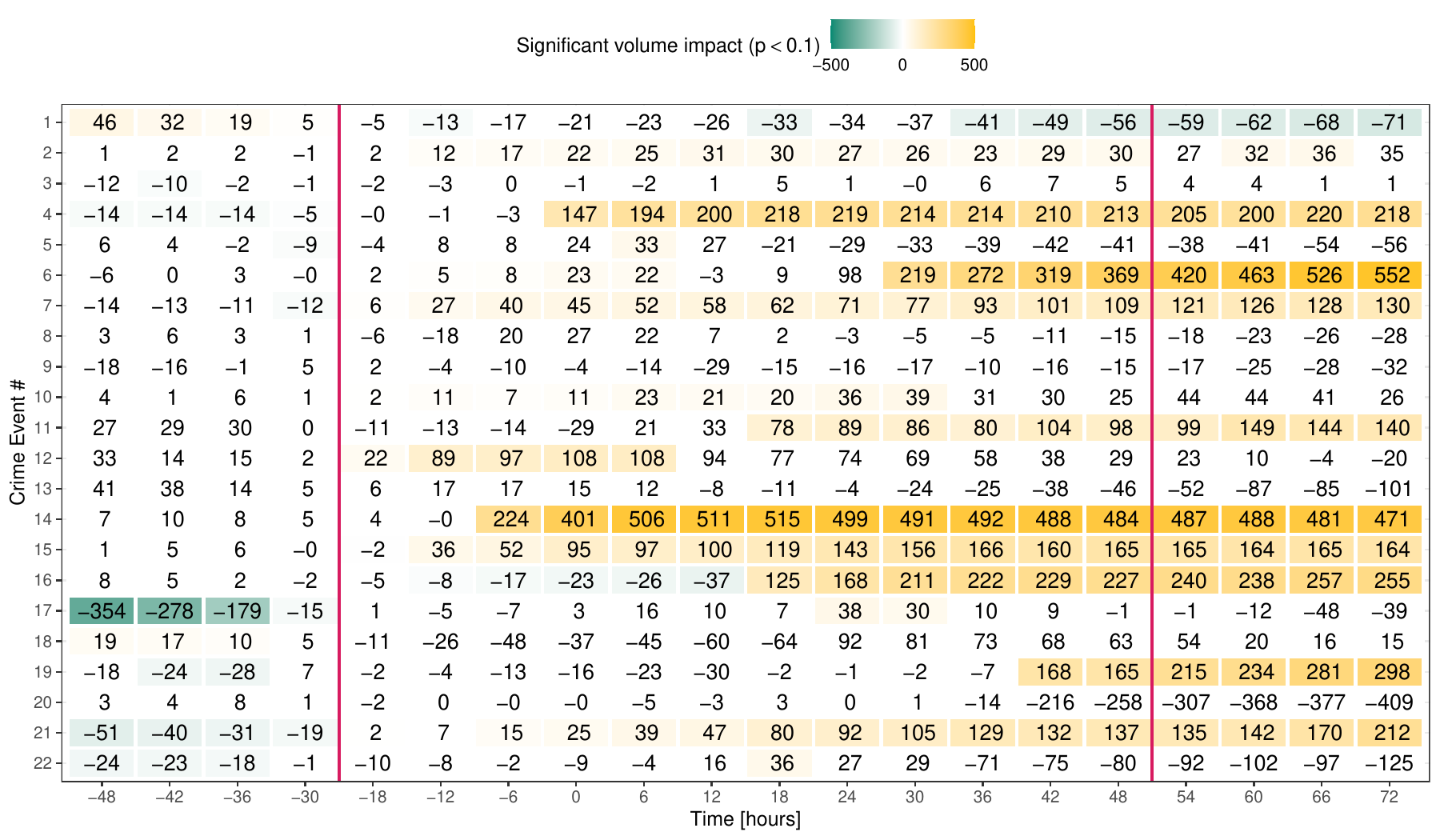}
	\caption{\textbf{Impact of DeFi Crime Events on DAO Governance Token Trading Volumes}: Each row represents the impact of a specific crime event on the associated DAO governance trading volume, with statistically significant values color-coded. \SigVolImpactPosN/ events exhibit a statistically significant increase in trading volume at least once within the event window $W^{E} = [-1, +2]$ days.}
	\label{fig:plDynDiDheatVol}
\end{figure}

The same method was applied to cumulative trading volumes for each crime event. The corresponding results are presented in Figure~\ref{fig:plDynDiDheatVol}. The dynamic effect analysis reveals a positive volume trend in \VolTrendPos/ crime events. However, \SigVolImpactPosN/ governance assets experienced a statistically significant increase ($p < 0.1$) in trading volume at least once within the event window $W^E$. Furthermore, a co-occurrence of a significant price decline and volume increase within the same time step $\Delta t$ is observed in \CoocSigImp/ out of \SigVolImpactPosN/ events. Notably, \SigVolThreeDigPosN/ governance assets exhibited trading volume increases exceeding $+100\%$. On average, the trading volume increased by $\text{\SigVolImpactPosPct/} \pm \text{\SigVolImpactPosSE/ \%}$, with a minimum of \SigVolImpactPosMin/\% and a maximum of \SigVolImpactPosMax/\%.

\emph{Overall, we found that \SigVolImpactPosNPct/\% of crime events had a statistically significant positive impact on the trading volume of the associated governance asset.}

\subsection{Impact of DeFi Crime Events on Market Capitalization}

Market capitalization is measured as the total number of assets multiplied by their price (as explained in Section~\ref{ssec:MarketCap}). To assess the indirect economic costs of crime events on market capitalization, we compare capitalization \emph{before} and \emph{after} each event.

Specifically, pre-event market capitalization is calculated using the price at the anchor point $t_a=-1$ day as the baseline, prior to the crime event $e$. Post-event market capitalization is estimated using the average of the most statistically significant prices. Significance levels are categorized as follows: $p < 0.001$ (***), $p < 0.01$ (**), $p < 0.05$ (*) $p < 0.1$ (\bulletSSS) or $p \geq 0.1$; the average calculation is restricted to the highest available significance class. The indirect economic impact was then measured as the difference between pre-event and post-event capitalization, but only for significant price declines.

In Table~\ref{tab:Implications}, we present (i) average price and trading volume estimates, with p-value classifications as described above; and (ii) the indirect economic impact, based on changes in market capitalization, alongside the direct economic impact (i.e., victims' losses) for each crime event, as reported by the prior study~\cite{CarpentierDesjardins2025} and their published data~\cite{carpentier-desjardins2024mapping}. We combine both direct and indirect impacts to calculate the total economic impact, expressing the indirect impact proportionally.

\emph{Overall, we found that across \SigPriceImpactNegN/ crime events, the total economic impact --- comprising both direct and indirect losses --- amounts to \TotalEcoImpact/.
Notably, indirect economic impacts far exceed direct losses, accounting for \IndirectEcoImpactPct/\% (\IndirectEcoImpact/) of the total impact and an average loss of \MCAvgSig/ million per DAO.}

\renewcommand{\arraystretch}{1.}

\begin{table}[!t]

	\begin{adjustwidth}{-0.cm}{-0.cm}

	\small
	\centering

\begin{tabular*}{1.0\textwidth}{@{\extracolsep{\fill}} 
		>{\arraybackslash\raggedleft}p{.5cm} 
		@{\hspace{0.15cm}}l@{\hspace{0.0cm}}
		@{\hspace{0.1cm}}r@{\hspace{0.15cm}}
		r| 
		S[table-format=-2.1, table-space-text-post=\,\bulletSSS, per-mode=symbol, table-align-text-pre=false, table-align-text-post=false, input-symbols=, input-open-uncertainty=, input-close-uncertainty=, detect-all]
		S[per-mode=symbol, table-align-text-pre=false, table-align-text-post=false, input-symbols=, input-open-uncertainty=, input-close-uncertainty=, detect-all]
		@{\hspace{0.15cm}}r@{\hspace{0.05cm}}
		@{\hspace{-0.1cm}}S[table-format=3.1, table-space-text-post=M, per-mode=symbol, table-align-text-pre=false, table-align-text-post=false, input-symbols=, input-open-uncertainty=, input-close-uncertainty=, detect-all]@{\hspace{0.05cm}}
		@{\hspace{0.15cm}}r@{\hspace{0.05cm}}
	}
	\toprule
	\textbf{\#} \vspace{1.5em} & 
	\parbox{1.25cm}{\textbf{Govner- nance \\Asset $a$}} & 
	\parbox{1.2cm}{\raggedleft\textbf{Direct \\Economic \\Impact }} &   
	\parbox{1.5cm}{\raggedleft\textbf{Market \\Capital- ization}} & 
	{\parbox{1cm}{\centering\textbf{Price \\Impact}}}  & 
	\parbox{1.cm}{\centering\textbf{Trading \\Volume \\Impact}} & 
	\multicolumn{2}{c}{\parbox{1.75cm}{\centering\textbf{Indirect Economic \\Impact}}} & 
	\parbox{1.5cm}{\raggedleft\textbf{Total \\Economic \\Impact}} \\ 
	&  & \text{[USD]} & \text{[USD]} & \text{[\%]} & \text{[\%]} & \text{[USD]} & \text{[\%]} & \text{[USD]} \\
	\midrule
	1 & CRV & 227k & 920.8M & -5.4\, \bulletSSS & -49.2\,* & 49.8M & (99.5\%) & 50.1M\\
	2 & PICKLE & 19.7M & 18.4M & 6.7\, \bulletSSS & 22.9\, *** & \text{-} &  & \text{-}\\
	3 & YFI & 129.5k & 446.2M & 1.7 & 1.2 & \text{-} &  & \text{-}\\
	4 & FARM & 24M & 75.8M & -59.3\,*** & 200.8\,*** & 45M & (65.2\%) & 69M\\
	5 & COMP & 89M & 1.4B & -1.5\,* & 32.8\, \bulletSSS & 20.5M & (18.8\%) & 109.5M\\
	6 & SUSHI & 15k & 210.3M & 36.0\,* & 270.2\,** & \text{-} &  & \text{-}\\
	7 & SUSHI & 105.2k & 1.3B & -12.7 & 58.4\,*** & \text{-} &  & \text{-}\\
	8 & YFI & 11M & 1B & -18.8\,* & 2.8 & 194.5M & (94.6\%) & 205.5M\\
	9 & CREAM & \text{-} & 1.1B & -2.5 & -12.1 & \text{-} &  & \text{-}\\
	10 & MASK & \text{-} & 1.7B & -7.2\, \bulletSSS & 23.4\,** & 120.7M &  & 120.7M\\
	11 & DAO & 7M & 615.3M & 11.2\, ** & 96.6\, ** & \text{-} &  & \text{-}\\
	12 & CREAM & 34M & 1.7B & -6.8\,*** & 88.6\,*** & 112.5M & (76.8\%) & 146.5M\\
	13 & DAO & 4M & 760.4M & 14.6\,*** & -3.9 & \text{-} &  & \text{-}\\
	14 & COMP & 80M & 3.3B & -9.8\,** & 458.5\,*** & 327.2M & (80.4\%) & 407.2M\\
	15 & COMP & 68.8M & 3.9B & -5.4\,*** & 112.2\,*** & 208.5M & (75.2\%) & 277.3M\\
	16 & NDX & 16M & 31.9M & -27.2\,** & 190.9\,*** & 8.7M & (35.2\%) & 24.7M\\
	17 & CREAM & 130M & 1.4B & -6.3\,*** & 30.0\,** & 90.7M & (41.1\%) & 220.7M\\
	18 & BADGER & 120M & 607.1M & -0.6 & 2.1 & \text{-} &  & \text{-}\\
	19 & FLURRY & 293k & 374.8k & -8.8\,** & 167.7\,*** & 33.1k & (10.2\%) & 326.1k\\
	20 & RBN & 351.1k & 195.1M & 3.0 & -21.5 & \text{-} &  & \text{-}\\
	21 & UNI & 8M & 5.4B & -5.7 & 129.4\,*** & \text{-} &  & \text{-}\\
	22 & CRV & 612k & 2.6B & -5.4\, * & 35.6\, \bulletSSS & 142.7M & (99.6\%) & 143.3M\\
	\midrule
	
	&  \multicolumn{2}{r}{ \textsuperscript{$\dagger$}\textbf{\DirectEcoImpact/} (613.2M)} \hspace{-1em} &  & & & \textbf{\IndirectEcoImpact/}  & \textbf{(\IndirectEcoImpactPct/\%)} & \textbf{\TotalEcoImpact/} \\
	\bottomrule
\end{tabular*}

	\caption{\textbf{Impact of DeFi Crime Events on DAO Governance Token Market Capitalization}:
	Each row represents a crime event, the associated governance asset, the direct economic impact and the market capitalization of the affected DAO. For each event, we report the price and trading volume impacts with statistical significance levels ($p < 0.001$ (***), $p < 0.01$ (**), $p < 0.05$ (*), $p < 0.1$ (\bulletSSS) or $p \geq 0.1$). The indirect economic impact is computed based on statistically significant price changes and compared with the direct losses incurred by victims. \textsuperscript{$\dagger$} Sum of the direct impact is considered only when a comparable value for the total economic impact is available.}
	\label{tab:Implications}

	\end{adjustwidth}
\end{table}


\section{Discussion}
\label{sec:discussion}

It is well established that DeFi services are prime targets for cybercriminals, often leading to substantial direct financial losses for victims.
However, unlike traditional economies, where cybercrime has known far-reaching economic consequences beyond direct costs imposed on targeted organizations and their customers, the indirect economic impact of DeFi crime on DAOs, a novel organizational model, remains largely unexplored. Therefore, in this study, we quantified the indirect economic impact of DeFi crime events by analyzing their effects on prices, trading volumes, and market capitalization of governance tokens associated with targeted DAOs. To achieve this, we curated a dataset containing price and volume data for \GovPricesN/ governance tokens to examine their responses to crime events. Between 2020 and 2022, we identified \UniqueGovAssetsN/ governance tokens linked to \DAOEventsAvailableN/ crime events.

\subsection{Key Takeaways}

Our study expands the existing body of knowledge on DeFi crime events affecting DAOs, with the following three key takeaways.

\paragraph*{Indirect costs of crime exceed victims' losses}
Previous studies have introduced the concept of the ``indirect cost of cybercrime''~\cite{Anderson2013,Anderson2019}. While assessing cyber harm to an organization is complex, the indirect economic impact on market capitalization is more straightforward to estimate.

We find that for \SigPriceImpactNegN/ out of \DAOEventsAvailableN/ crime events, the DAOs' governance tokens lose a market capitalization of \IndirectEcoImpact/ USD. This accounts for \IndirectEcoImpactPct/\% of the total loss, meaning that the indirect economic impact surpasses direct crime costs. Notably, this figure is derived from a conservative estimate, relying only on statistically significant values from the DiD model.

This market reaction does not primarily affect the DAO itself; rather, it impacts governance token holders, akin to shareholders, who ultimately bear the value loss. Although cryptoasset prices can recover, and the impact may diminish over time, declines can also persist beyond the event window. Hence, the non-negligible indirect costs of cybercrime should not be overlooked, as this study shows.
DAOs may experience additional indirect consequences of cybercrime, such as decaying usage of the decentralized application or other network-related and social effects. Quantifying these additional costs would be a promising research avenue for further understanding this novel organizations.

\paragraph*{DAOs experience stronger price declines from crime than traditional organizations}

Our study shows that affected DAOs' governance tokens experience an average price decline of $\SigPriceImpactNegPct/ \pm \SigPriceImpactNegSE/\%$ with a range from
\SigPriceImpactNegMin/\% to \SigPriceImpactNegMax/\%. By comparison, previous meta-studies on traditional organizations report significantly smaller stock price declines: an average decline of 3.5\%, ranging from -0.25\% to -10\%~\cite{Ali2021a}; declines between -0.8\% and -2.3\%~\cite{Woods2021}; and an average decline of -0.953\%~\cite{Ebrahimi2022}. These results indicate that DAOs, as a novel and still developing organizational structure, exhibit stronger market reactions to cybercrime than traditional organizations.

Furthermore, we observe a statistically significant price decline in \SigPriceImpactNegN/ and a trading volume increase in \SigVolImpactPosN/ out of \DAOEventsAvailableN/ crime events. Price declines and volume increases occur simultaneously in \CoocSigImp/ events - i.e., both effects within the same time step $\Delta t$. Price and volume may respond jointly to information flows~\cite{Chen2001}, such as the announcement of a crime event. In this context, the observed co-movement in \CoocSigImp/ events could indicate that these crime events were high profile and/or that the information about the crime was disseminated quickly, amplifying market reactions. Comparing these significant crime events with those that did not yield any market movement could be the topic of a follow-up study.
The dataset published [\textit{{link accessible upon publication}}] can act as a starting point for such studies.

Also, note that cumulative trading volume does not differentiate between buying and selling activity. Consequently, market participants may react differently to crime events: some might lose trust and sell tokens, while others may see an opportunity to acquire assets at a lower price. This aligns with findings in the cryptocurrency literature, which documents varied market reactions, sometimes even in opposite directions~\cite{Brown2020,Milunovich2022a}.
Given these findings, further studies should investigate how token holders react to such events, including what factors make them buy and/or sell. Moreover, given that recent studies have shown that DAOs face power concentration and unexercised voting rights~\cite{Dotan2023a, Fritsch2022a}, further studies should also look into how crime events change governance dynamics, including potential shifts in power structures within DAOs.
These would provide a more granular understanding of the complex behavioral and market dynamics that shape these novel governance structures.

\paragraph*{Correlated counterfactual assets enable strong statistical inference}

In event studies, market-based models help relate event-driven reactions to a baseline by comparing it with the overall trend, and consequently, allowing for stronger causal inferences~\cite{HuntingtonKlein}.
However, unlike traditional markets that have established benchmarks such as the market or sector-specific indices, cryptocurrencies do not have yet similar reliable reference points, as previous research has noted~\cite{Abramova2021}.

To address this limitation, we use counterfactual governance assets as a control units. Our proposed procedure successfully identifies several highly correlated counterfactuals for many governance tokens, largely satisfying the DiD parallel trends assumption, as observed in the reference window before the event.
We leverage actual governance tokens rather than synthetic assets, as the latter would compromise the validity of our approach towards causal inference. Notably, the same governance asset may serve different counterfactual roles across various crime events, which we attribute to the high volatility of the DeFi market and the relatively recent emergence of these assets and liquidity pools. This dynamic environment makes traditional methods, such as placebo tests, challenging to apply effectively.
For instance, \textit{Sushiswap} DAO, a major DEX on Ethereum, saw its market capitalization surge from $210.3$M to $1.3$B USD between 2020-11-30 ($\tau_6$) and 2021-01-27 ($\tau_7$), driven by a sharp price increase in both ETH and SUSHI.

Overall, we identified on average $\PriceCorAvgN/$ counterfactual assets for the price and $\VolCorAvgN/$ for the volume time series, both highly correlated with the analyzed governance token.
Rather than relying on market indices, counterfactual assets allow us to adapt the DiD framework. We believe this approach could benefit other studies, as many event studies on cryptoassets lack a market model or a suitable market index. Hence, future research can further refine this methodology, extending its application to other decentralized markets where traditional benchmarks may fall short.

\subsection{Limitations and Future Work}

Our study has certain limitations and opens new directions for future research.

First, we focus on the DeFi hype period between 2020 and 2022. During this phase, new tokens and services emerged, making it particularly relevant to study market reactions within this still immature and vulnerable ecosystem. However, extending the analysis beyond this period could provide further insights into how the decentralized financial ecosystem, crime, and market reactions evolve over time. Future research should therefore investigate more recent events to examine whether market behavior and resilience to negative external events are changing.

Second, our analysis is restricted to on-chain price and trading volume data from Uniswap V2 on Ethereum.
This approach considers settled transactions, which are recorded on the blockchain, making them immutable and reconstructible.
However, it would also be valuable to study the impact of crime events on centralized exchanges (CEXs), which handle higher trading volumes and liquidity.
Future research could compare different DEXs across multiple blockchains with their centralized counterparts to better understand markets.

Finally, our study does not fully reflect the complex relationship between characteristics of DeFi criminal events and market reactions.
Particularly, understanding why certain events have a larger influence on DAOs than others is left out of scope.
Is it due to the crime type, the targeted governance asset, or a combination of both?
Is it due to the specifics of a crime or the broadcasting in the news?
These open questions may unveil insights about the dynamic expectations and responses of market participants.


\section{Conclusion}
\label{sec:conclusion}

In this study, we examined market reactions to DeFi crime events affecting governance assets and found that such events are associated with significant price declines and increased trading volumes. The indirect costs of these crimes amount to \IndirectEcoImpact/ (\IndirectEcoImpactPct/\%) of \TotalEcoImpact/ of the total costs. Given that indirect costs may include additional factors, such as cybersecurity expenses~\cite{Anderson2013, Anderson2019}, the economic harm inflicted on DAOs is substantial. Therefore, proactively strengthening security on DAOs and DeFi platforms is a worthwhile investment to mitigate potential risks before becoming a victim. Beyond the analytical findings, our study adapts the DiD model and introduces a reproducible measurement framework to quantify the indirect economic impact of crime on publicly traded cryptoassets. The dataset, compiled for this study and made available online, also provides a valuable resource for further research. Through these contributions, we provide insights into the market dynamics and resilience of DeFi technologies, ultimately advancing economic and computational research on this evolving and novel financial ecosystem.

\newpage

\bibliography{literature}

\begin{thebibliography}{10}

\bibitem{Abhishta2019}
Abhishta Abhishta, Reinoud Joosten, Sergey Dragomiretskiy, and Lambert J.~M.
  Nieuwenhuis.
\newblock Impact of successful {DDoS} attacks on a major crypto-currency
  exchange.
\newblock In {\em 2019 27th Euromicro International Conference on Parallel,
  Distributed and Network-Based Processing (PDP)}, pages 379--384, February
  2019.
\newblock ISSN: 2377-5750.
\newblock URL: \url{https://ieeexplore.ieee.org/document/8671642}, \href
  {https://doi.org/10.1109/EMPDP.2019.8671642}
  {\path{doi:10.1109/EMPDP.2019.8671642}}.

\bibitem{Abramova2021}
Svetlana Abramova and Rainer Böhme.
\newblock Out of the dark: The effect of law enforcement actions on
  cryptocurrency market prices.
\newblock In {\em 2021 APWG Symposium on Electronic Crime Research (eCrime)},
  pages 1--11, December 2021.
\newblock ISSN: 2159-1245.
\newblock URL:
  \url{https://ieeexplore.ieee.org/document/9738787/keywords#keywords}, \href
  {https://doi.org/10.1109/eCrime54498.2021.9738787}
  {\path{doi:10.1109/eCrime54498.2021.9738787}}.

\bibitem{Adams2020}
Hayden Adams, Noah Zinsmeister, and Dan Robinson.
\newblock Uniswap v2 core.
\newblock {\em URL: https://uniswap.org/whitepaper.pdf}, 5:10--15, 2020.

\bibitem{Agrafiotis2018}
Ioannis Agrafiotis, Jason R.~C. Nurse, Michael Goldsmith, Sadie Creese, and
  David Upton.
\newblock A taxonomy of cyber-harms: Defining the impacts of cyber-attacks and
  understanding how they propagate.
\newblock {\em Journal of Cybersecurity}, 4(1):tyy006, January 2018.
\newblock \href {https://doi.org/10.1093/cybsec/tyy006}
  {\path{doi:10.1093/cybsec/tyy006}}.

\bibitem{Ali2021a}
Syed Emad~Azhar Ali, Fong-Woon Lai, P.~D.~D. Dominic, Nicholas~James Brown,
  Paul~Benjamin Lowry, and Rao~Faizan Ali.
\newblock Stock market reactions to favorable and unfavorable information
  security events: A systematic literature review.
\newblock {\em Computers \& Security}, 110:102451, November 2021.
\newblock URL:
  \url{https://www.sciencedirect.com/science/article/pii/S0167404821002753},
  \href {https://doi.org/10.1016/j.cose.2021.102451}
  {\path{doi:10.1016/j.cose.2021.102451}}.

\bibitem{Ali2021}
Syed Emad~Azhar Ali, Fong-Woon Lai, Rohail Hassan, and Muhammad~Kashif Shad.
\newblock The long-run impact of information security breach announcements on
  investors' confidence: The context of efficient market hypothesis.
\newblock {\em Sustainability}, 13(3):1066, 2021.
\newblock URL: \url{https://www.mdpi.com/2071-1050/13/3/1066}.

\bibitem{Anderson2019}
Ross Anderson, Chris Barton, Rainer Böhme, Richard Clayton, Carlos Ganán, Tom
  Grasso, Michael Levi, Tyler Moore, and Marie Vasek.
\newblock Measuring the changing cost of cybercrime.
\newblock 2019.
\newblock URL: \url{https://orca.cardiff.ac.uk/id/eprint/122684/}.

\bibitem{Anderson2013}
Ross Anderson, Chris Barton, Rainer Böhme, Richard Clayton, Michel J.~G. van
  Eeten, Michael Levi, Tyler Moore, and Stefan Savage.
\newblock Measuring the cost of cybercrime.
\newblock pages 265--300. Springer, Berlin, Heidelberg, 2013.
\newblock \href {https://doi.org/10.1007/978-3-642-39498-0_12}
  {\path{doi:10.1007/978-3-642-39498-0_12}}.

\bibitem{Angeris2024}
Guillermo Angeris, Tarun Chitra, Theo Diamandis, Kshitij Kulkarni, and Alex
  Evans.
\newblock The geometry of constant function market makers.
\newblock In {\em Proceedings of the 25th ACM Conference on Economics and
  Computation}, EC '24, page 732, New York, NY, USA, December 2024. Association
  for Computing Machinery.
\newblock \href {https://doi.org/10.1145/3670865.3673500}
  {\path{doi:10.1145/3670865.3673500}}.

\bibitem{Ante2023}
Lennart Ante.
\newblock How {Elon Musk}'s {Twitter} activity moves cryptocurrency markets.
\newblock {\em Technological Forecasting and Social Change}, 186:122112,
  January 2023.
\newblock URL:
  \url{https://www.sciencedirect.com/science/article/pii/S0040162522006333},
  \href {https://doi.org/10.1016/j.techfore.2022.122112}
  {\path{doi:10.1016/j.techfore.2022.122112}}.

\bibitem{Appel2023a}
Ian Appel and Jillian Grennan.
\newblock Decentralized governance and digital asset prices.
\newblock Technical Report 4367209, Rochester, NY, February 2023.
\newblock URL: \url{https://papers.ssrn.com/abstract=4367209}, \href
  {https://doi.org/10.2139/ssrn.4367209} {\path{doi:10.2139/ssrn.4367209}}.

\bibitem{Auer2018}
Raphael Auer and Stijn Claessens.
\newblock Regulating cryptocurrencies: Assessing market reactions.
\newblock Technical Report 3288097, Rochester, NY, September 2018.
\newblock URL: \url{https://papers.ssrn.com/abstract=3288097}.

\bibitem{Auer2024}
Raphael Auer, Bernhard Haslhofer, Stefan Kitzler, Pietro Saggese, and Friedhelm
  Victor.
\newblock The technology of decentralized finance ({DeFi}).
\newblock {\em Digital Finance}, 6(1):55--95, March 2024.
\newblock \href {https://doi.org/10.1007/s42521-023-00088-8}
  {\path{doi:10.1007/s42521-023-00088-8}}.

\bibitem{Barbereau2022}
Tom Barbereau, Reilly Smethurst, Orestis Papageorgiou, Johannes Sedlmeir, and
  Gilbert Fridgen.
\newblock Decentralised finance's unregulated governance: Minority rule in the
  digital wild west.
\newblock Technical Report 4001891, Rochester, NY, January 2022.
\newblock URL: \url{https://papers.ssrn.com/abstract=4001891}, \href
  {https://doi.org/10.2139/ssrn.4001891} {\path{doi:10.2139/ssrn.4001891}}.

\bibitem{Brown2020}
Michael~Scott Brown and Barry Douglass.
\newblock An event study of the effects of cryptocurrency thefts on
  cryptocurrency prices.
\newblock In {\em 2020 Spring Simulation Conference (SpringSim)}, pages 1--12,
  May 2020.
\newblock \href {https://doi.org/10.22360/SpringSim.2020.CSE.001}
  {\path{doi:10.22360/SpringSim.2020.CSE.001}}.

\bibitem{Brown1985}
Stephen~J. Brown and Jerold~B. Warner.
\newblock Using daily stock returns: The case of event studies.
\newblock {\em Journal of Financial Economics}, 14(1):3--31, March 1985.
\newblock URL:
  \url{https://www.sciencedirect.com/science/article/pii/0304405X8590042X},
  \href {https://doi.org/10.1016/0304-405X(85)90042-X}
  {\path{doi:10.1016/0304-405X(85)90042-X}}.

\bibitem{Buterin2014}
Vitalik Buterin.
\newblock A next-generation smart contract and decentralized application
  platform.
\newblock {\em white paper}, 3(37):2--1, 2014.
\newblock URL:
  \url{https://www.weusecoins.com/assets/pdf/library/Ethereum_white_paper-a_next_generation_smart_contract_and_decentralized_application_platform-vitalik-buterin.pdf}.

\bibitem{Campbell1996}
Cynthia~J. Campbell.
\newblock Measuring abnormal daily trading volumes for samples of {NYSE}/{ASE}
  and {NASDAQ} securities using parametric and nonparametric test statistics.
\newblock {\em Review of Quantitative Finance and Accounting}, 6(3), 1996.
\newblock URL: \url{https://works.bepress.com/cynthia_campbell/6/}.

\bibitem{Campbell2003}
Katherine Campbell, Lawrence~A. Gordon, Martin~P. Loeb, and Lei Zhou.
\newblock The economic cost of publicly announced information security
  breaches: Empirical evidence from the stock market.
\newblock {\em Journal of Computer Security}, 11(3):431--448, January 2003.
\newblock URL:
  \url{https://content.iospress.com/articles/journal-of-computer-security/jcs192},
  \href {https://doi.org/10.3233/JCS-2003-11308}
  {\path{doi:10.3233/JCS-2003-11308}}.

\bibitem{carpentier-desjardins2024mapping}
C.~Carpentier-Desjardins, M.~Paquet-Clouston, S.~Kitzler, and B.~Haslhofer.
\newblock Mapping the {DeFi} crime landscape: An evidence-based picture, 2024.
\newblock \href {https://doi.org/10.5281/zenodo.14047933}
  {\path{doi:10.5281/zenodo.14047933}}.

\bibitem{CarpentierDesjardins2025}
Catherine Carpentier-Desjardins, Masarah Paquet-Clouston, Stefan Kitzler, and
  Bernhard Haslhofer.
\newblock Mapping the {DeFi} crime landscape: An evidence-based picture.
\newblock {\em Journal of Cybersecurity}, 11(1):tyae029, 2025.
\newblock URL:
  \url{https://academic.oup.com/cybersecurity/article-abstract/11/1/tyae029/7962044}.

\bibitem{Chen2001}
Gong-meng Chen, Michael Firth, and Oliver~M. Rui.
\newblock The dynamic relation between stock returns, trading volume, and
  volatility.
\newblock {\em Financial Review}, 36(3):153--174, 2001.
\newblock URL:
  \url{https://onlinelibrary.wiley.com/doi/abs/10.1111/j.1540-6288.2001.tb00024.x},
  \href {https://doi.org/10.1111/j.1540-6288.2001.tb00024.x}
  {\path{doi:10.1111/j.1540-6288.2001.tb00024.x}}.

\bibitem{Chen2023}
Yu-Lun Chen, Yung~Ting Chang, and J.~Jimmy Yang.
\newblock Cryptocurrency hacking incidents and the price dynamics of {Bitcoin}
  spot and futures.
\newblock {\em Finance Research Letters}, 55:103955, July 2023.
\newblock URL:
  \url{https://www.sciencedirect.com/science/article/pii/S1544612323003276},
  \href {https://doi.org/10.1016/j.frl.2023.103955}
  {\path{doi:10.1016/j.frl.2023.103955}}.

\bibitem{Dotan2023a}
Maya Dotan, Aviv Yaish, Hsin-Chu Yin, Eytan Tsytkin, and Aviv Zohar.
\newblock The vulnerable nature of decentralized governance in {DeFi}.
\newblock In {\em Proceedings of the 2023 Workshop on Decentralized Finance and
  Security}, DeFi '23, pages 25--31, New York, NY, USA, November 2023.
  Association for Computing Machinery.
\newblock URL: \url{https://dl.acm.org/doi/10.1145/3605768.3623539}, \href
  {https://doi.org/10.1145/3605768.3623539}
  {\path{doi:10.1145/3605768.3623539}}.

\bibitem{Ebrahimi2022}
Sepideh Ebrahimi and Kamran Eshghi.
\newblock A meta-analysis of the factors influencing the impact of security
  breach announcements on stock returns of firms.
\newblock {\em Electronic Markets}, 32(4):2357--2380, December 2022.
\newblock \href {https://doi.org/10.1007/s12525-022-00550-2}
  {\path{doi:10.1007/s12525-022-00550-2}}.

\bibitem{Ershov2020EC}
Daniel Ershov and Matthew Mitchell.
\newblock The effects of influencer advertising disclosure regulations:
  Evidence from {Instagram}.
\newblock In {\em Proceedings of the 21st ACM Conference on Economics and
  Computation}, EC '20, pages 73--74, New York, NY, USA, 2020. Association for
  Computing Machinery.
\newblock \href {https://doi.org/10.1145/3391403.3399477}
  {\path{doi:10.1145/3391403.3399477}}.

\bibitem{Fama1970}
Eugene~F. Fama.
\newblock Efficient capital markets: A review of theory and empirical work.
\newblock {\em The Journal of Finance}, 25(2):383--417, 1970.
\newblock URL: \url{https://www.jstor.org/stable/2325486}, \href
  {https://doi.org/10.2307/2325486} {\path{doi:10.2307/2325486}}.

\bibitem{Feichtinger2024}
Rainer Feichtinger, Robin Fritsch, Lioba Heimbach, Yann Vonlanthen, and Roger
  Wattenhofer.
\newblock {SoK}: Attacks on {DAOs}, August 2024.
\newblock arXiv:2406.15071 [cs].
\newblock URL: \url{http://arxiv.org/abs/2406.15071}, \href
  {https://doi.org/10.48550/arXiv.2406.15071}
  {\path{doi:10.48550/arXiv.2406.15071}}.

\bibitem{Fritsch2022a}
Robin Fritsch, Marino Müller, and Roger Wattenhofer.
\newblock Analyzing voting power in decentralized governance: Who controls
  {DAOs}?, April 2022.
\newblock arXiv:2204.01176 [cs].
\newblock URL: \url{http://arxiv.org/abs/2204.01176}, \href
  {https://doi.org/10.48550/arXiv.2204.01176}
  {\path{doi:10.48550/arXiv.2204.01176}}.

\bibitem{Harvey2020}
Campbell~R. Harvey, Ashwin Ramachandran, and Joey Santoro.
\newblock {DeFi} and the future of finance.
\newblock Technical report, 2020.
\newblock URL:
  \url{https://papers.ssrn.com/sol3/papers.cfm?abstract_id=3711777}.

\bibitem{heimbach2021behavior}
Lioba Heimbach, Ye~Wang, and Roger Wattenhofer.
\newblock Behavior of liquidity providers in decentralized exchanges.
\newblock {\em arXiv:2105.13822}, 2021.

\bibitem{HuntingtonKlein}
Nick Huntington-Klein.
\newblock {\em The Effect: An Introduction to Research Design and Causality}.
\newblock Chapman and Hall/CRC, 2021.

\bibitem{Iglewicz1993}
Boris Iglewicz and David~C. Hoaglin.
\newblock {\em Volume 16: How to detect and handle outliers}.
\newblock Quality Press, January 1993.

\bibitem{Kiayias2023a}
Aggelos Kiayias and Philip Lazos.
\newblock {SoK}: Blockchain governance.
\newblock In {\em Proceedings of the 4th ACM Conference on Advances in
  Financial Technologies}, AFT '22, pages 61--73, New York, NY, USA, July 2023.
  Association for Computing Machinery.
\newblock URL: \url{https://dl.acm.org/doi/10.1145/3558535.3559794}, \href
  {https://doi.org/10.1145/3558535.3559794}
  {\path{doi:10.1145/3558535.3559794}}.

\bibitem{Kiayias2023}
Aggelos Kiayias and Philip Lazos.
\newblock {SoK}: Blockchain governance, January 2023.
\newblock arXiv:2201.07188 [cs].
\newblock URL: \url{http://arxiv.org/abs/2201.07188}, \href
  {https://doi.org/10.48550/arXiv.2201.07188}
  {\path{doi:10.48550/arXiv.2201.07188}}.

\bibitem{Kitzler2023a}
Stefan Kitzler, Stefano Balietti, Pietro Saggese, Bernhard Haslhofer, and
  Markus Strohmaier.
\newblock The governance of decentralized autonomous organizations: A study of
  contributors' influence, networks, and shifts in voting power, September
  2023.
\newblock arXiv:2309.14232 [cs].
\newblock URL: \url{http://arxiv.org/abs/2309.14232}, \href
  {https://doi.org/10.48550/arXiv.2309.14232}
  {\path{doi:10.48550/arXiv.2309.14232}}.

\bibitem{Laturnus2023}
Valerie Laturnus.
\newblock The economics of decentralized autonomous organizations, January
  2023.
\newblock URL: \url{https://papers.ssrn.com/abstract=4320196}, \href
  {https://doi.org/10.2139/ssrn.4320196} {\path{doi:10.2139/ssrn.4320196}}.

\bibitem{Lehar2022}
Alfred Lehar and Christine~A. Parlour.
\newblock Systemic fragility in decentralized markets.
\newblock Technical Report 4164833, Rochester, NY, July 2022.
\newblock URL: \url{https://papers.ssrn.com/abstract=4164833}, \href
  {https://doi.org/10.2139/ssrn.4164833} {\path{doi:10.2139/ssrn.4164833}}.

\bibitem{Li2024}
Zhe Li.
\newblock {\em The impact of blockchain security breaches}.
\newblock {PhD} thesis, University of British Columbia, 2024.
\newblock URL:
  \url{https://open.library.ubc.ca/soa/cIRcle/collections/ubctheses/24/items/1.0440990}.

\bibitem{Liu2021}
Lu~Liu, Sicong Zhou, Huawei Huang, and Zibin Zheng.
\newblock From technology to society: An overview of blockchain-based {DAO}.
\newblock {\em IEEE Open Journal of the Computer Society}, 2:204--215, 2021.
\newblock \href {https://doi.org/10.1109/OJCS.2021.3072661}
  {\path{doi:10.1109/OJCS.2021.3072661}}.

\bibitem{Milunovich2022a}
George Milunovich and Seung~Ah Lee.
\newblock Measuring the impact of digital exchange cyberattacks on {Bitcoin}
  returns.
\newblock {\em Economics Letters}, 221:110893, December 2022.
\newblock URL:
  \url{https://www.sciencedirect.com/science/article/pii/S0165176522003676},
  \href {https://doi.org/10.1016/j.econlet.2022.110893}
  {\path{doi:10.1016/j.econlet.2022.110893}}.

\bibitem{Saggese2025}
Pietro Saggese, Michael Fröwis, Stefan Kitzler, Bernhard Haslhofer, and
  Raphael Auer.
\newblock Towards verifiability of total value locked ({TVL}) in decentralized
  finance.
\newblock 2025.
\newblock URL: \url{https://www.bis.org/publ/work1268.htm}.

\bibitem{Spanos2016}
Georgios Spanos and Lefteris Angelis.
\newblock The impact of information security events to the stock market: A
  systematic literature review.
\newblock {\em Computers \& Security}, 58:216--229, May 2016.
\newblock URL:
  \url{https://www.sciencedirect.com/science/article/pii/S0167404816300013},
  \href {https://doi.org/10.1016/j.cose.2015.12.006}
  {\path{doi:10.1016/j.cose.2015.12.006}}.

\bibitem{Tan2023}
Joshua~Z. Tan, Tara Merk, Sarah Hubbard, Eliza~R. Oak, Joni Pirovich, Ellie
  Rennie, Rolf Hoefer, Michael Zargham, Jason Potts, Chris Berg, Reuben
  Youngblom, Primavera De~Filippi, Seth Frey, Jeff Strnad, Morshed Mannan,
  Kelsie Nabben, Silke~Noa Elrifai, Jake Hartnell, Benjamin~Mako Hill, Alexia
  Maddox, Woojin Lim, Tobin South, Ari Juels, and Dan Boneh.
\newblock Open problems in {DAOs}, October 2023.
\newblock arXiv:2310.19201 [cs].
\newblock URL: \url{http://arxiv.org/abs/2310.19201}, \href
  {https://doi.org/10.48550/arXiv.2310.19201}
  {\path{doi:10.48550/arXiv.2310.19201}}.

\bibitem{Tosun2021}
Onur~Kemal Tosun.
\newblock Cyber-attacks and stock market activity.
\newblock {\em International Review of Financial Analysis}, 76:101795, July
  2021.
\newblock URL:
  \url{https://www.sciencedirect.com/science/article/pii/S1057521921001319},
  \href {https://doi.org/10.1016/j.irfa.2021.101795}
  {\path{doi:10.1016/j.irfa.2021.101795}}.

\bibitem{Werner2022}
Sam~M. Werner, Daniel Perez, Lewis Gudgeon, Ariah Klages-Mundt, Dominik Harz,
  and William~J. Knottenbelt.
\newblock {SoK}: Decentralized finance ({DeFi}), September 2022.
\newblock arXiv:2101.08778 [cs, econ, q-fin].
\newblock URL: \url{http://arxiv.org/abs/2101.08778}, \href
  {https://doi.org/10.48550/arXiv.2101.08778}
  {\path{doi:10.48550/arXiv.2101.08778}}.

\bibitem{Woods2021}
Daniel~W. Woods and Rainer Böhme.
\newblock {SoK}: Quantifying cyber risk.
\newblock In {\em 2021 IEEE Symposium on Security and Privacy (SP)}, pages
  211--228, May 2021.
\newblock ISSN: 2375-1207.
\newblock URL: \url{https://ieeexplore.ieee.org/document/9519490}, \href
  {https://doi.org/10.1109/SP40001.2021.00053}
  {\path{doi:10.1109/SP40001.2021.00053}}.

\bibitem{Xu2022a}
Jiahua Xu, Krzysztof Paruch, Simon Cousaert, and Yebo Feng.
\newblock {SoK}: Decentralized exchanges ({DEX}) with automated market maker
  ({AMM}) protocols.
\newblock {\em arXiv:2103.12732 [cs, q-fin]}, January 2022.
\newblock arXiv:2103.12732.
\newblock URL: \url{http://arxiv.org/abs/2103.12732}.

\bibitem{Emrah2022}
Emrah Öget.
\newblock The effect of positive and negative events on cryptocurrency prices.
\newblock {\em Ekonomi Politika ve Finans Araştırmaları Dergisi},
  7(1):16--31, 2022.
\newblock URL: \url{https://dergipark.org.tr/en/pub/epfad/issue/68438/1011204}.

\end{thebibliography}

\appendix

\clearpage

\section{Appendix}

\subsection{Matching Entries}\label{par:matching}

To merge the crime database with Ethereum data, we use an automated matching approach to link crime actors with cryptoassets. Specifically, we compare the actors' names in the crime database to two cryptoasset name features: the Coingecko identifier (id) and token name.
To ensure consistency, we first clean all text entries by removing special characters. Next, we tokenize each entry into three-character word chunks and use Jaccard similarity to measure the overlap between the actor names and cryptoasset name features.
We establish matching thresholds (0.8 for id, 0.7 for name) by manually verifying that governance assets of DAOs with the highest TVL\footnote{\url{https://defillama.com/chain/Ethereum}} are included. Additionally, if a cryptoasset name contains DeFi-related keywords (e.g., 'DAO' or 'network'), which may reduce similarity scores, we apply a lower threshold (0.5 for name matching with keywords).
Finally, we identify cryptoassets that exceed at least one threshold and match them to the corresponding actors in the crime database.

\subsection{DEX Cleaning}\label{ssec:Clearning}

To ensure our analysis is based on mature data, we exclude DEX time series that do not meet minimum criteria for continuous trading.
In particular, in the early stages of a DEX pool, insufficient liquidity or low trading activity can make price reaction analyses unreliable.
Other indicators of immaturity can include short existence periods or highly irregular time gaps between data points.
Therefore, we only include assets that meet specific liquidity criteria.
First, the price series must span at least 1 million blocks, which corresponds to approximately $10^6 \times 15s = 416$ days, ensuring coverage of at least one annual cycle.
Second, we require an average of at least one data point per day. Given an estimated Ethereum block time of 15 seconds, we impose a threshold where the average interval between two price points must not exceed 6,000 blocks ($ 6000 \times 15s = 25$  hours).

\subsection{Counterfactual Assets}\label{ssec:AppendixCounterfactuals}

\begin{figure}
	\centering
	\includegraphics[width=0.8\columnwidth]{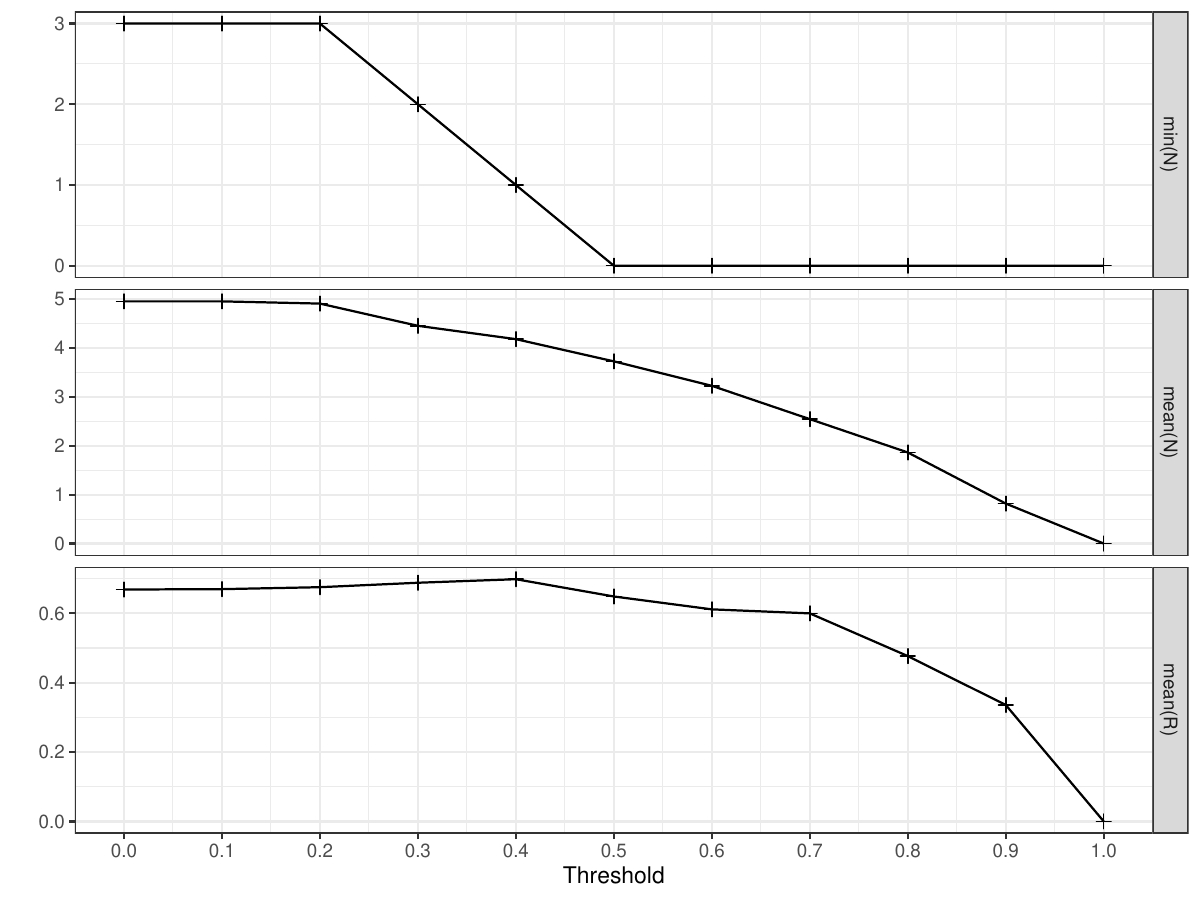}
	\caption{\textbf{Counterfactual Assets Threshold:}
		Varying the minimum thresholds for the counterfactual asset procedure results in different numbers of assets and correlations with the targeted governance asset for each crime event. The figure illustrates both the minimum and mean number of assets, as well as the average correlation across events. The threshold of $t = .4$ strikes a good balance, ensuring at least one asset per event while maintaining an average correlation at the optimum.
	}
	\label{fig:pl_price_sim}
\end{figure}

The identification procedure in Section~\ref{sec:Method} introduces a minimum threshold for the correlation between governance assets.
For robustness, we evaluate different thresholds and show in Figure~\ref{fig:pl_price_sim} how different thresholds lead to varying sets in terms of size and correlation.
The threshold of $t = 0.4$ strikes the balance between average correlation and ensuring at least one asset per event.

\begin{figure}
	\includegraphics[width=1\columnwidth]{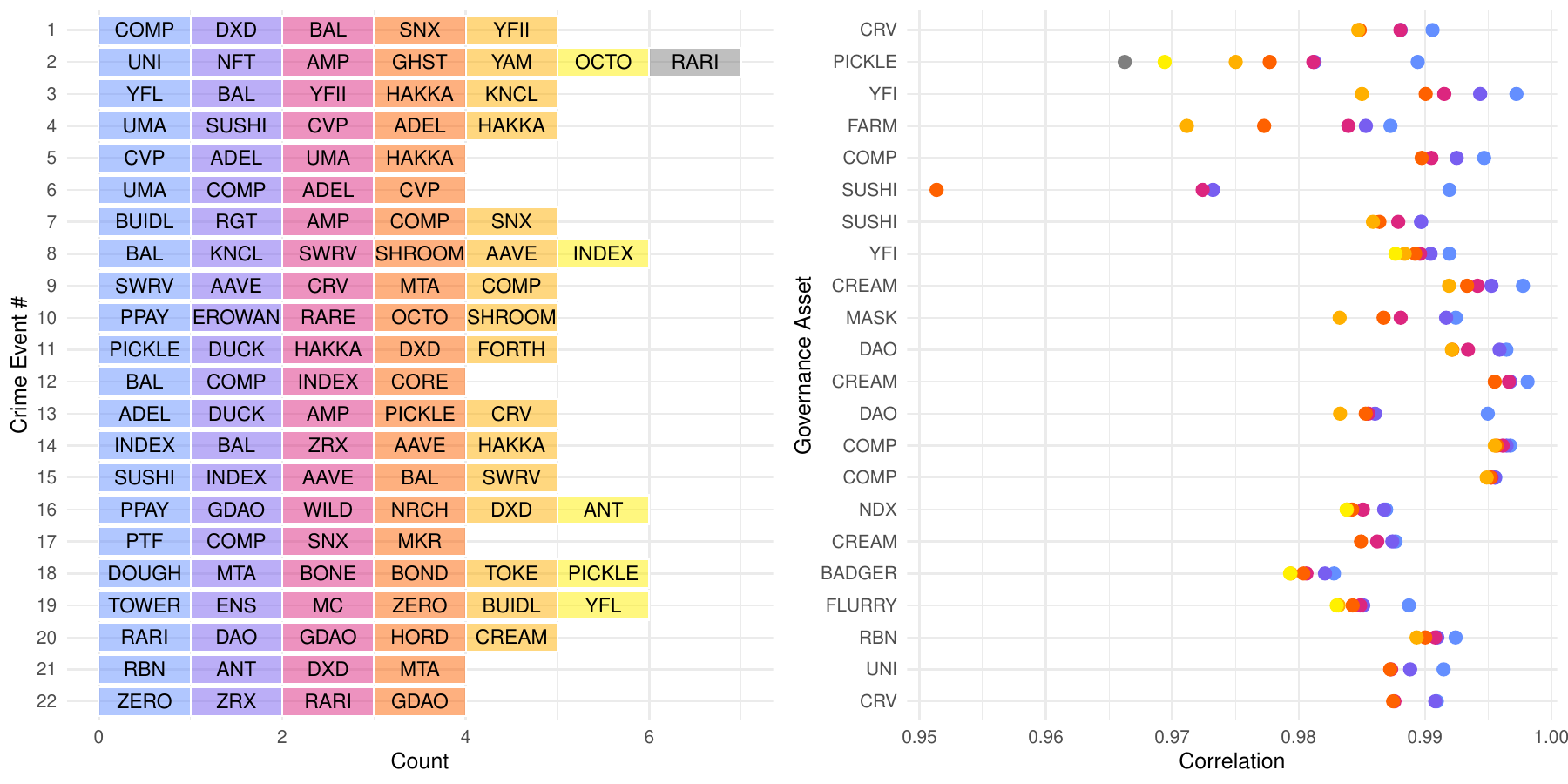}
	\caption{\textbf{Counterfactual Assets:} For each crime event $e$, we identified counterfactual assets (left) for the targeted governance asset $a$, based on the correlation (on the right) in historical cumulated trading volume during the reference period $W^{L}$. }
	\label{fig:pl_cor_vol}
\end{figure}

We apply the identification procedure to the counterfactual assets based on price and volume.
In Figure~\ref{fig:pl_cor_vol}, we present the supplementary identified counterfactual assets and their correlations for each crime event, based on the cumulative volume.

\subsection{Dynamic DiD Coefficient-plot}\label{ssec:AppendixDiD}

\begin{figure}
	\centering
	\includegraphics[width=0.8\columnwidth]{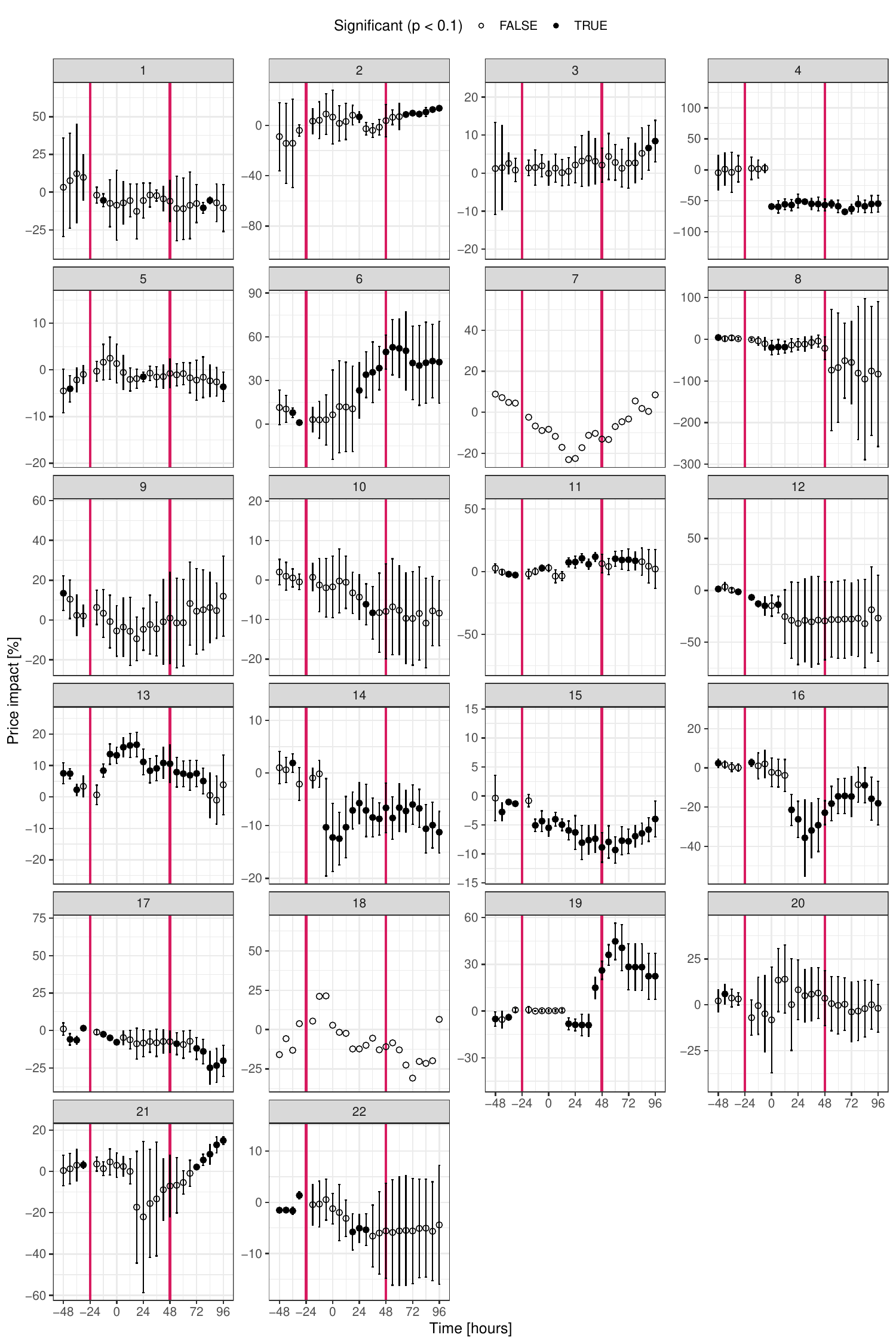}
	\caption{\textbf{Price Impact for each Crime Events} We present the regression coefficients from the governance price analysis along with the 90\% confidence interval, separately for each crime event. The red lines indicate the event window $W^E$, with the first red line also serving as the anchor point, which serves as a reference for the regression.}\label{fig:coefPlotCAll}
\end{figure}

\begin{figure}
	\centering
	\includegraphics[width=0.8\columnwidth]{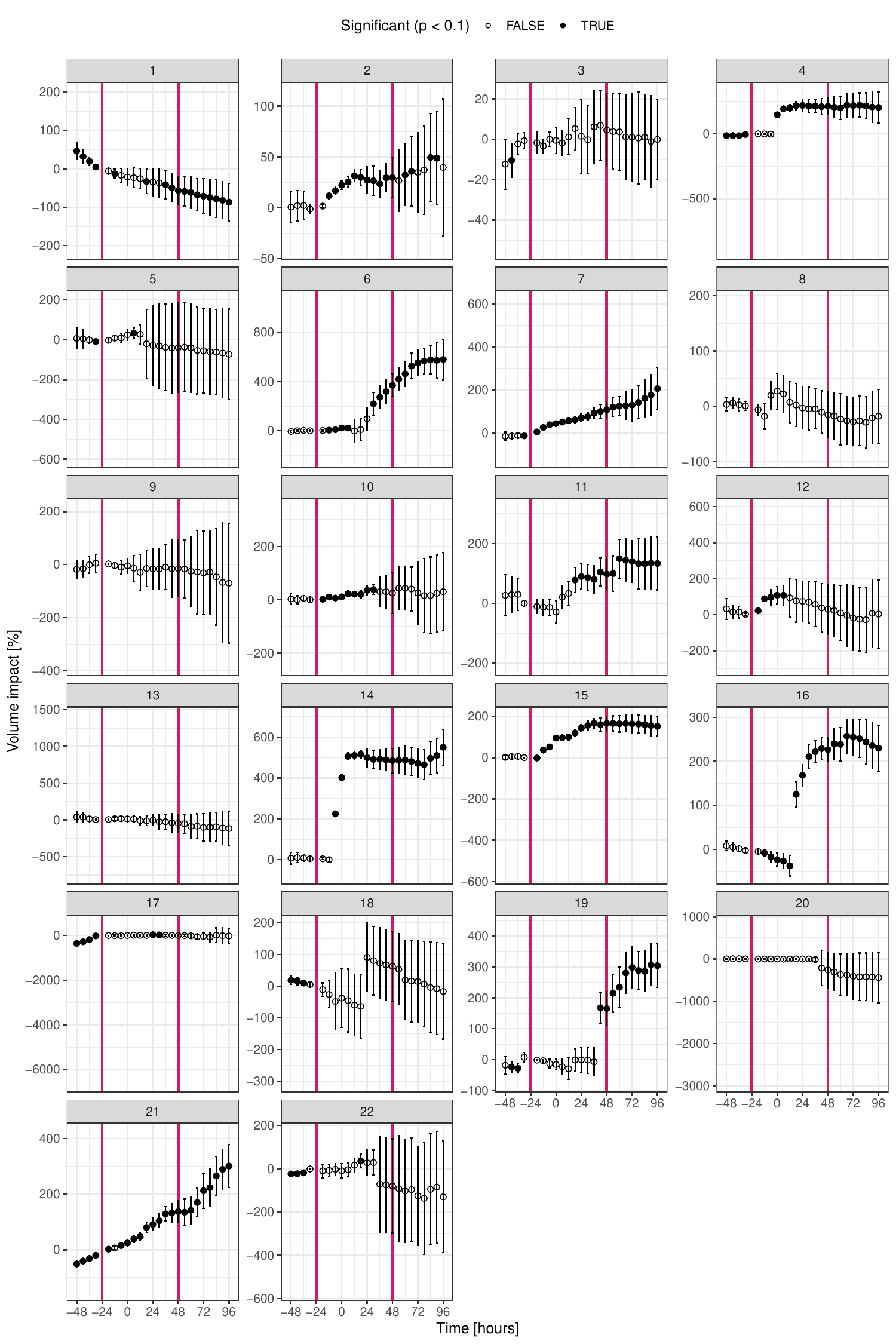}
	\caption{\textbf{Volume Impact for each Crime Events} For each crime event, we present the regression coefficients from the governance trading volume analysis along with the 90\% confidence interval. The figure captures the event window $W^E$, indicated by red vertical lines.}\label{fig:coefPlotVolCAll}
\end{figure}

We apply the Difference-in-Differences (DiD) method to each crime event and their identified counterfactual assets.
In addition to the estimated coefficients in Section~\ref{sec:results},
we present the temporal evolution of each crime event separately,
including bars for the $90\%$ confidence interval and black-colored dots as indicators of significance ($p<0.1$).
Figure~\ref{fig:coefPlotCAll} shows the DiD for the price analysis, while Figure~\ref{fig:coefPlotVolCAll} presents the DiD for the cumulative trading volume.

\newpage

\end{document}